# SAMRI: Segment Any MRI

Zhao Wang[1], Wei Dai[1], Thuy Thanh Dao[1], Steffen Bollmann[1,2], Hongfu Sun[3], Craig Engstrom[4], Shekhar S. Chandra[1]

*[1]School of Electrical Engineering and Computer Science, The University of Queensland, Australia*

*[2]Queensland Digital Health Centre, University of Queensland, Australia*

*[3]School of Engineering, College of Engineering, Science and Environment, University of Newcastle, Australia*

*[4]School of Human Movement and Nutrition Sciences, The University of Queensland, Australia*

## Abstract

**Summary:** SAMRI is an MRI-specialized adaptation of the Segment Anything Model achieving superior whole-body MRI segmentation, particularly for small and clinically critical structures, through box and point prompts for rapid annotation.

**Purpose:** Existing SAM adaptations treat MRI as a generic modality, overlooking variable tissue contrast, intensity inhomogeneity, and clinically important small structures. We propose an MRI-specialized foundation model with strong whole-body segmentation and zero-shot generalization for direct use on any MRI annotation task.

**Methods:** SAMRI fine-tunes only the mask decoder of SAM (ViT-B/16), keeping encoders frozen to preserve pretrained representations and eliminate redundant passes—reducing training time by 94%, trainable parameters by 96%, and FLOPs by ~99% versus full-model retraining. Training used 1.1 million 2D slice–mask pairs from 30 datasets spanning 47 targets, T1/T2/FLAIR/DWI contrasts, and whole-body anatomy, with focal–Dice loss and bounding-box (with optional point) prompts. Sizes were stratified by mask area (small: <0.5%; medium: 0.5–3.5%; large: >3.5%), and significance assessed by the Wilcoxon signed-rank test.

**Results:** SAMRI with box+point prompts achieved mean DSC 0.87 ± 0.11 across 47 targets, outperforming MedSAM (0.74 ± 0.24) by 17.6% ($p < 0.05$), with largest gains for small (+42.4%) and medium (+26.9%) structures. On six zero-shot datasets, SAMRI achieved mean DSC 0.85, outperforming baselines. Inference requires only ~4.5 GB VRAM through an interactive interface on standard hardware.

**Conclusion:** Decoder-only fine-tuning on a large, MRI-specific corpus delivers superior whole-body segmentation with strong zero-shot generalization, particularly for small and clinically salient structures. Public code, pretrained models, and an interactive interface make SAMRI deployable for MRI segmentation research and clinical workflows.

## Introduction

Magnetic resonance imaging (MRI) provides exquisite soft-tissue visualization, making it well suited for clinical tasks such as diagnosis, disease monitoring, and treatment planning [1]. However, MRI segmentation—delineating anatomical structures or pathological regions from MRI scans—is labor-intensive, requires well-trained experts, and remains prone to inter-observer variability [2]. These challenges arise from MRI's variable tissue contrast (e.g., T1, T2, FLAIR, DWI), intensity inhomogeneity, and acquisition protocols [3,4,5], and are compounded by the prevalence of small, clinically critical structures—such as microbleeds (2–5 mm), small metastases (<10 mm), thin cartilage (<3 mm), and early-stage lesions—that may occupy only 10–100 pixels even in high-resolution scans [6,7,8,9,10].

Early automated segmentation methods—atlas-based registration, statistical shape modeling, and intensity-driven clustering—leveraged anatomical priors or voxel-level intensity statistics to reduce manual effort [11,12,13,14]. Atlas-based registration ensures anatomically consistent labels but is computationally expensive and results degrade with registration errors. Statistical shape models impose structural plausibility yet fail when pathological shapes deviate from trained priors. Intensity-driven clustering methods are simple and training-free but are sensitive to MRI's intensity nonuniformity and overlapping tissue contrasts. These limitations motivated the shift toward convolutional neural networks (CNNs), which learn hierarchical representations directly from data to achieve more accurate and generalizable segmentation.

Convolutional neural networks have since become the dominant paradigm in medical image segmentation due to their hierarchical feature extraction and strong representational power [15,16]. Representative architectures such as U-Net and its numerous variants (e.g., VNet, 3D U-Net, Attention U-Net, nnU-Net)—along with other CNN-based designs like CAN3D—have achieved state-of-the-art performance in both 2D and volumetric segmentation [17,18,19,20,21,22,23,24,25,26]. Despite their success, CNN-based models require task-specific training and must be fully retrained when segmentation targets change, constraining scalability across scanners, contrasts, and anatomical regions [27,28,29,30,31,32]. These limitations motivate interest in foundation models capable of broad generalization across tasks and domains [33].

The Segment Anything Model (SAM), pretrained on 11 million natural images with 1.1 billion masks, introduces a prompt-driven segmentation framework supporting zero-shot inference via user-provided points or bounding boxes [34]. SAM's architecture consists of a Vision Transformer image encoder, a prompt encoder, and a lightweight mask decoder, enabling interactive segmentation without task-specific retraining. Despite strong generalization on natural images, SAM applied directly to MRI underperforms due to fundamental domain differences. Standard object-size benchmarks (e.g., COCO bounding-box thresholds [35]) predominantly feature large, natural-scene objects, whereas MRI exhibits subtle tissue boundaries, pronounced foreground–background class imbalance, and a high proportion of small or irregularly shaped structures [36,37,38,39]—features largely absent from SAM's pretraining distribution [34]. This mismatch motivates MRI-aware size definitions, introduced in the Methods section.

Several strategies have been proposed to adapt SAM for medical image segmentation (Supplementary Table S2). **Full fine-tuning** approaches such as MedSAM [40] update all SAM parameters on multi-modality medical datasets; however, dilution across non-MRI modalities limits MRI-specific performance [41], and the high computational cost makes full-model fine-tuning on MRI data infeasible in most research settings. **Encoder distillation** approaches (EMedSAM [42]) and **adapter-based methods** (MA-SAM [43], Med-SA [44]) reduce trainable parameters, but are typically validated on narrow anatomical targets and still require repeated online feature extraction during each training epoch—making overall training substantially slower than their reduced parameter count suggests. **Prompt-enhancement methods** (DeSAM [45], MedLSAM [46]) retain SAM's original weights but rely on pretrained natural-image features, constraining adaptation to MRI's contrast heterogeneity and small-structure challenges. **Prompt-encoder modification** approaches such as MCP-MedSAM [47] introduce learnable modality and content prompts into the prompt encoder, achieving efficient multi-modal training on a single GPU within one day; however, training across 11 imaging modalities degrades MRI-specific fidelity, as reflected in its lower mean DSC ($0.73 \pm 0.22$) on our MRI benchmark. Critically, all strategies that modify the encoder—through fine-tuning, adapters, or distillation—risk degrading its general visual features, reducing cross-contrast generalizability across diverse MRI protocols.

To address these limitations, we present SAMRI, an MRI-specialized SAM adaptation that achieves strong segmentation performance—particularly for small, clinically critical structures—while enabling practical deployment through a decoder-only training strategy (Fig. 1). SAMRI precomputes image embeddings using SAM's frozen encoder, then trains only the mask decoder on a curated corpus of 36 MRI datasets covering 47 anatomical and pathological tasks, over 10 MRI contrasts, and 1.1 million slice–mask pairs spanning whole-body anatomy. Freezing the encoder preserves its pretrained visual representations, eliminates repeated online encoding during training, and substantially reduces hardware requirements—enabling training on a single commercial GPU. Our main contributions are:

(i) **Robust generalization:** mean DSC 0.87 across 47 tasks and 0.85 in zero-shot evaluation on unseen anatomies and contrasts, surpassing MedSAM (0.74) by 17.6% and demonstrating the breadth of cross-protocol generalizability.

(ii) **Superior performance on small-to-medium structures:** mean DSC 0.84 (small) and 0.85 (medium) versus MedSAM 0.59 and 0.67, yielding gains of up to +42.4% and +26.9%, respectively—addressing a critical gap for clinically important MRI targets.

(iii) **Efficient two-stage training:** freezing all encoders and training only the mask decoder with a combined focal–Dice loss reduces training time by 94%, trainable parameters by 96%, and training FLOPs by approximately 99% compared with full-model retraining, making MRI-specific foundation model adaptation feasible on standard hardware.

(iv) **Practical deployment:** a complete training-to-inference pipeline with a user-friendly graphical user interface (GUI) for interactive segmentation on standard hardware, together with public code and pretrained models to support reproducibility and direct adoption by the MRI research community.

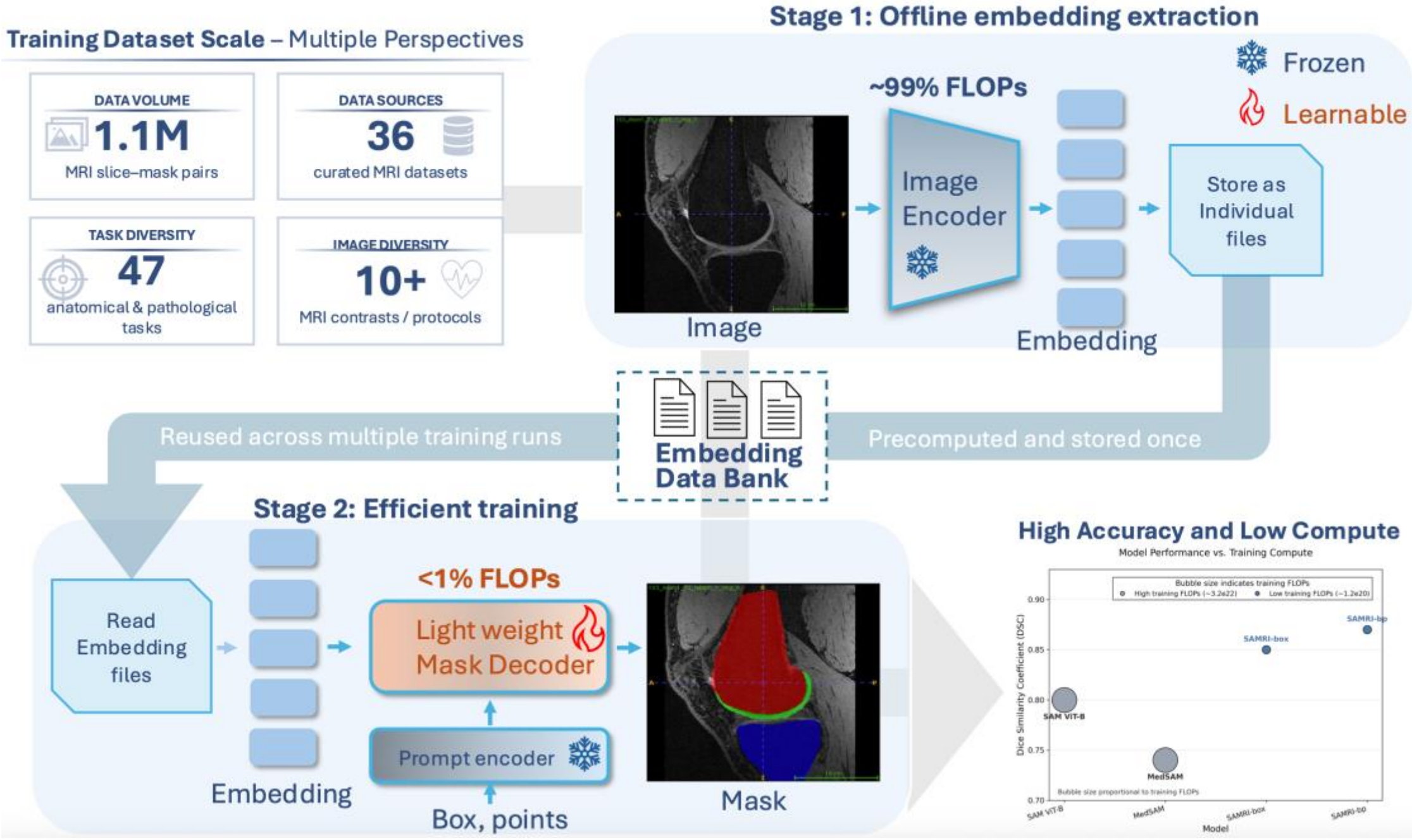

Fig. 1 The SAMRI training framework. A large-scale MRI dataset (1.1 M slice–mask pairs from 36 datasets, 47 tasks, and 10+ contrasts) is used with a two-stage strategy—**Stage 1**: one-time offline embedding extraction with a frozen image encoder, and **Stage 2**: efficient training that reuses stored embeddings to optimize only a lightweight mask decoder with prompt inputs, achieving high accuracy with substantially reduced computational cost compared to full-model training.

## Methods

Motivated by the limitations of existing SAM-based approaches for MRI segmentation, SAMRI adopts a two-stage training framework tailored to MRI data. This section describes the training architecture and procedure, large-scale MRI dataset construction and preprocessing, loss function, training protocol, object size stratification strategy, baseline models, and statistical analysis approach.

### SAMRI Two-Stage Training Procedure

SAMRI builds on SAM (ViT-B/16) by freezing the heavyweight image encoder and training only the lightweight mask decoder. Reusing SAM's pretrained encoder—responsible for most of SAM's computational cost during both training and inference (Supplementary Table S5)—allows us to precompute embeddings once (Stage 1, Fig. 1) and fine-tune only the decoder for domain shift (Stage 2, Fig. 1), drastically reducing training cost. Empirically, in our setting, SAM ViT-H, the largest SAM variant, did not yield measurably better MRI segmentation than SAM ViT-B (Supplementary Fig. S9), suggesting that ViT-B features are already sufficient for MRI textures and contrasts; a similar observation was reported by others [38]. By concentrating learning in the decoder, SAMRI adapts specifically to MRI characteristics while preserving computational efficiency and improving segmentation accuracy. Removing the encoder from the online training loop eliminates redundant forward passes, yielding an approximate 88% reduction in total training time at N = 100 epochs (theoretical analysis in Supplementary Appendix S1); the empirically observed reduction over our full 200-epoch training run is 94%.

## Dataset Collection and Preprocessing

**Diversified Data Sources.** We assembled 36 MRI datasets incorporating multi-organ/body region imaging spanning examinations of both healthy and pathological tissues: 21 inherited from MedSAM and 15 new MRI datasets added with additional tasks and contrasts. In the final corpus, about 72% of images come from the new datasets and 28% from MedSAM's collection (Supplementary Fig. S1), broadening anatomical and contrast diversity and supporting stronger zero-shot generalization. Sources include Kaggle, OAI, TCIA, and MICCAI segmentation challenges. The collection covers 47 segmentation tasks across multiple MRI contrasts (T1, T2, FLAIR, DWI) and anatomical regions (brain, spine, knee, prostate, heart, abdomen). In total, we curated ~1.1 million expert-annotated 2D image–mask pairs. A visual summary of the data sources is presented in Fig. 2, with a detailed breakdown provided in Table S1 in Supplementary.

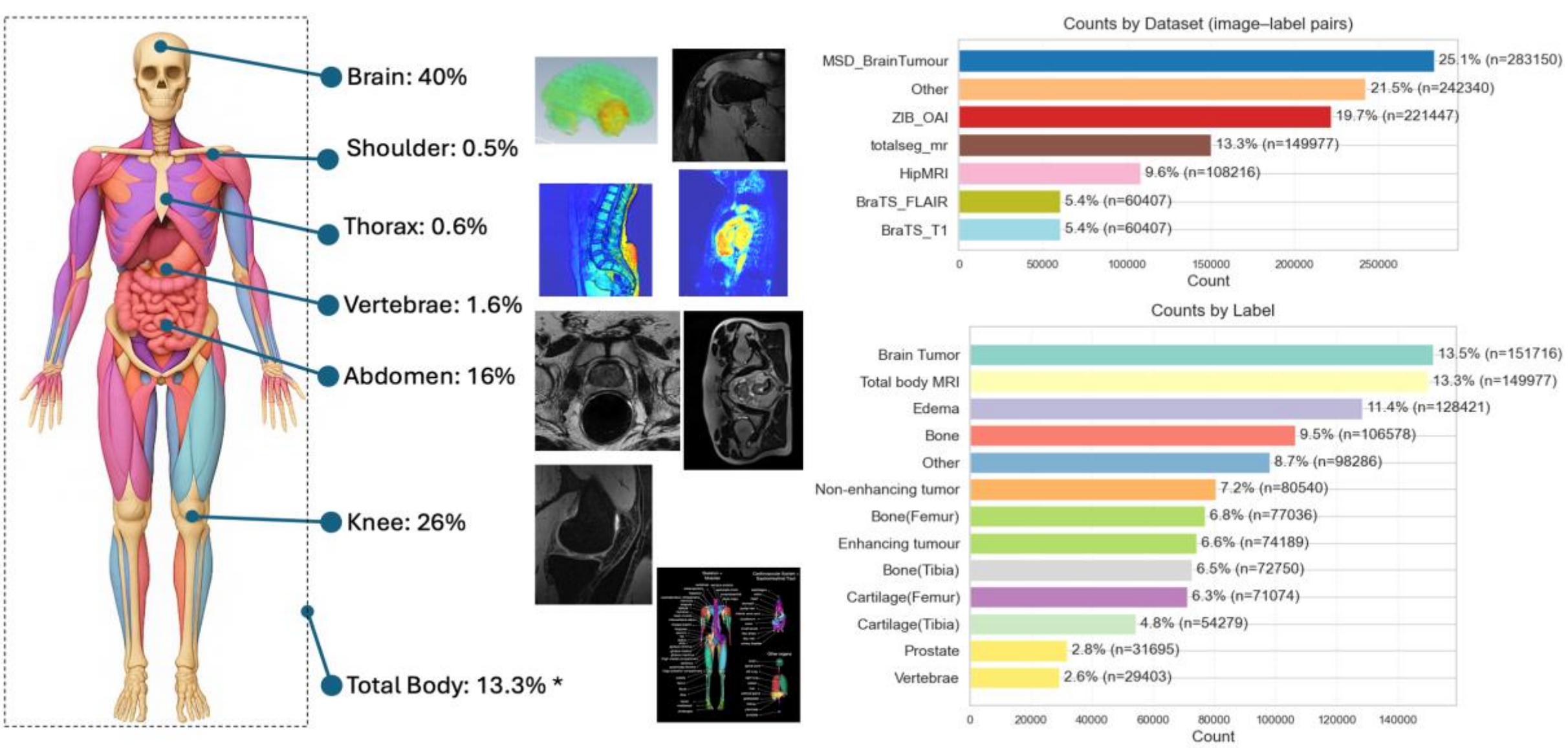

Fig. 2 Composition of the SAMRI training corpus—anatomical coverage and dataset/task mix. **Left:** anatomical coverage of the 1.1 million MRI image-mask pairs used to train SAMRI, summarized by body region (Brain 40%, Knee 26%, Abdomen 16%, Vertebrae 2.6%, Shoulder 0.5%, Thorax 0.2%, and a Whole-body set 13.3%*). **Right:** distribution by dataset (top) and by task/structure (bottom). The corpus draws broadly from public MR datasets (e.g., MSD Brain Tumour, ZIB OAI, BraTS, HipMRI, and others), and spans a diverse set of targets including brain tumour subregions, whole-body organs, cartilage (femur/tibia), bone, edema, prostate, and vertebrae. This breadth across anatomies and sequences provides the variability needed for robust training and strong zero-shot generalization. * "Total Body" denotes whole-body, multi-organ MRI annotations from TotalSeg-MR for general purpose segmentation tasks.

**Preprocessing Pipeline.** The collected MRI data were in various formats (NIfTI, MATLAB, DICOM, NRRD, MetaImage Header) and included both 2D and 3D acquisitions. We standardized all data by:

- Converting all 3D MRI volumes into 2D slices in NIfTI format along the anatomical axis with the lowest dimension, ensuring each slice captures maximal structural details, and discarding all masks containing fewer than 10 labeled pixels. This preprocessing protocol follows MedSAM's approach to ensure a fair comparison.
- Performing patient-level splits (80%:10%:10% for train, validation, and test) to prevent data leakage.
- For the training set only, discarding masks with fewer than 10% of the largest slice's mask area, as such small masks typically arise from peripheral slices with incomplete anatomical coverage or annotation artifacts, which impair training stability (see Supplementary Fig. S11). Validation and test sets are left unfiltered to preserve evaluation integrity.
- Normalizing grayscale MRI slices to [0, 255] integer range and replicating across three channels to match SAM's RGB input requirements.

After pre-processing, the images were passed through the frozen SAM encoder to generate visual embeddings, which were computed once and stored on the Embedding Data Bank (Hard disk or data cluster) for reuse during decoder training (Fig. 1).

**Sampling Strategy.** To address dataset-level imbalance (ranging from 56 to 283K samples per dataset), we applied per-epoch resampling so that every dataset contributes a minimum of 5,000 samples. Specifically, small datasets are sampled with replacement until they reach this quota, while large datasets are subsampled without replacement—yielding a near-uniform dataset contribution per epoch. This raises the sampling probability of anatomies that occur primarily in small datasets, preventing gradient updates from being dominated by the largest corpora and thereby ensuring broader anatomical coverage during training.

**Training and zero-shot datasets.** We partitioned 36 MRI datasets into two groups (Supplementary Table S1): 30 datasets were used for the standard train, validation, and testing workflow; six were held out entirely for zero-shot evaluation. The 30-set pool was selected to maximize anatomical and contrast diversity (vendors, contrasts, resolutions), improving robustness and reducing overfitting. For the zero-shot benchmark, five datasets mirror MedSAM's zero-shot suite to enable like-for-like comparison, while the sixth—MSK shoulder—was added to probe cross-anatomy transfer. Importantly, all six datasets are unseen for all evaluated models (SAMRI, SAM, MedSAM, and MCP-MedSAM), ensuring a fair comparison. Because all training cartilage data are from the knee, performance on shoulder cartilage in MSK shoulder dataset assesses whether the model exploits structural similarity rather than anatomy-specific shortcuts. This design isolates generalization, prevents leakage between training and evaluation, and directly measures robustness to dataset and anatomical variations.

## Loss Function

To train SAMRI effectively across diverse MRI segmentation tasks, we utilized a loss function that combines the strengths of both focal loss [48] and Dice loss [49]. The final loss $\mathcal{L}_{SAMRI}$ is defined as:

$$\mathcal{L}_{\text{SAMRI}} = 20 \cdot \mathcal{L}_{\text{Focal}} + \mathcal{L}_{\text{Dice}}$$

This formulation is particularly effective for small object segmentation, as focal loss prevents rare pixels (often corresponding to small structures) from being overwhelmed during training, while Dice loss ensures accurate boundary delineation, which is crucial when objects are present in sparse, low-pixel-count regions. $\mathcal{L}_{\text{Focal}}$ and $\mathcal{L}_{\text{Dice}}$ are defined in Supplementary Appendix S2.

## Training protocol

The SAMRI model was initialized with the pre-trained SAM using the ViT-Base backbone. The prompt encoder was kept frozen, as it is already well-suited for processing bounding box prompts. To simulate interactive segmentation, we generate synthetic box prompts from the ground truth segmentation masks by extracting the tightest bounding box around each individual object in the image. To improve robustness and generalization, we apply random spatial shifts to the box coordinates by adding or subtracting up to 20 pixels during training. To retain its generalization capability in feature abstraction, the image encoder was also frozen during training. In contrast, the light-weight mask decoder was trained. This selective training strategy enables our efficient training procedure while maintaining SAM's robust feature representations. Training was conducted using the AdamW optimizer [31] with $\beta_1$ = 0.9, $\beta_2$ = 0.999, an initial learning rate of 1e-5, and weight decay of 0.1. We used a global batch size of 1024 without data augmentation, given the scale and diversity of the 30 training datasets. Training was performed for 200 epochs on 8 AMD MI300X GPUs (each with 192 GB VRAM).

## Object Size Definition and Stratification

As motivated in the Introduction, COCO's bounding-box-based size thresholds are ill-suited for MRI, where thin or elongated structures can span large bounding boxes while occupying minimal mask area. To address this, we define object size by mask area rather than bounding-box area, normalized by image resolution. Specifically, we bin samples by percentage of image area (mask pixels / image pixels), with cut-offs at 0.5% and 3.5% to form small, medium, and large categories (comparable to COCO's cut-offs at 0.3% and 3%). Size bins are detailed in Table 1.

## Baseline Models

We evaluate SAMRI against three representative SAM-based 2D models using DSC, HD, and MSD as evaluation metrics (details in Supplementary Appendix S3). SAMRI and SAM use identical pre/post-processing; MedSAM and MCP-MedSAM follow their published pipelines. All methods receive identical ground-truth-derived bounding-box prompts with no model-specific tuning.

- **SAM (ViT-B)**. Official pretrained weights used in zero-shot evaluation, serving as the canonical foundation baseline without adaptation.

- **MedSAM**. A medical adaptation of SAM trained on ~1.5 million images across modalities (MRI, CT, ultrasound, X-ray), representing the current best general-purpose medical segmentation model based on SAM.

- **MCP-MedSAM**. A prompt-encoder modification approach that replaces SAM's prompt encoder with learnable modality embeddings (derived from CLIP text features) and content prompts (from a CNN branch applied to the bounding-box crop), trained across 11 imaging modalities on a single GPU within one day; included as a representative prompt-encoder modification baseline.

**Inclusion criteria and rationale.** We include only models with publicly available pretrained checkpoints suitable for MRI-wide, zero-shot evaluation. Within this scope:

- **ViT-H vs ViT-B**. On our development set, SAM ViT-H offered no consistent advantage over ViT-B in zero-shot MRI while being much heavier (Supplementary Fig. S9); we therefore adopt ViT-B as the canonical SAM baseline.
- **Adapter-based variants**. Most are trained primarily on CT or single-task MRI; for models with publicly released checkpoints (Med-SA, SAM-Med2D, SAMed), we conducted additional experiments confirming that a single pretrained checkpoint does not generalize across our full range of MRI contrasts, anatomies, and tasks (see Supplementary Fig. S10 and Table S2)
- **Prompt-enhanced methods**. These typically reuse the original SAM weights (e.g., DeSAM, MedLSAM in Supplementary Table S2); differences largely reflect prompting heuristics rather than a distinct foundation model, so their intrinsic capacity aligns with the SAM baseline.
- **MCP-MedSAM**. Included to represent prompt-encoder modification strategies; its multi-modal training scope and publicly released checkpoint make it a direct benchmark for assessing whether modality-agnostic training preserves MRI-specific fidelity.

## Statistical Analysis

Statistical significance was assessed using the Wilcoxon signed-rank test for paired comparisons between models across all 47 segmentation targets. This non-parametric test was chosen as the performance metrics were not normally distributed across diverse anatomical structures. We report p-values with $p < 0.05$ indicating significant differences.

# Results

This section presents a comprehensive evaluation of SAMRI in terms of segmentation accuracy, object size–stratified performance, training computational efficiency, and generalization. We first report overall quantitative results across 47 MRI targets and compare SAMRI with SAM ViT-B, MedSAM, and MCP-MedSAM. We then analyze performance gains for small anatomical structures, provide qualitative visual comparisons, examine zero-shot generalization to unseen datasets, and assess training behavior and practical deployment considerations.

## Segmentation Performance

We evaluated SAMRI—including box-only ($\text{SAMRI}_{\text{box}}$) and box+point ($\text{SAMRI}_{\text{bp}}$) prompt variants—against SAM ViT-B, MedSAM, and MCP-MedSAM (Figs. 3 and 4). Across 47 anatomical and pathological targets from 36 datasets, $\text{SAMRI}_{\text{bp}}$ achieved the best overall performance, with the highest mean DSC ($0.87 \pm 0.11$) and lowest boundary errors (HD: $5.36 \pm 4.07$ px; MSD: $1.30 \pm 0.76$ px). $\text{SAMRI}_{\text{box}}$ also performed well (DSC: $0.85 \pm 0.15$;

HD: 5.95 ± 6.21 px; MSD: 1.42 ± 1.10 px), with $\text{SAMRI}_{bp}$ showing modest but consistent advantages across most targets. Both SAMRI variants significantly outperformed MedSAM, MCP-MedSAM, and SAM ViT-B across all metrics (Wilcoxon signed-rank test, $p < 0.05$). Full per-target and per-dataset results are provided in Supplementary Tables S3–S4 and Figs. S2–S8.

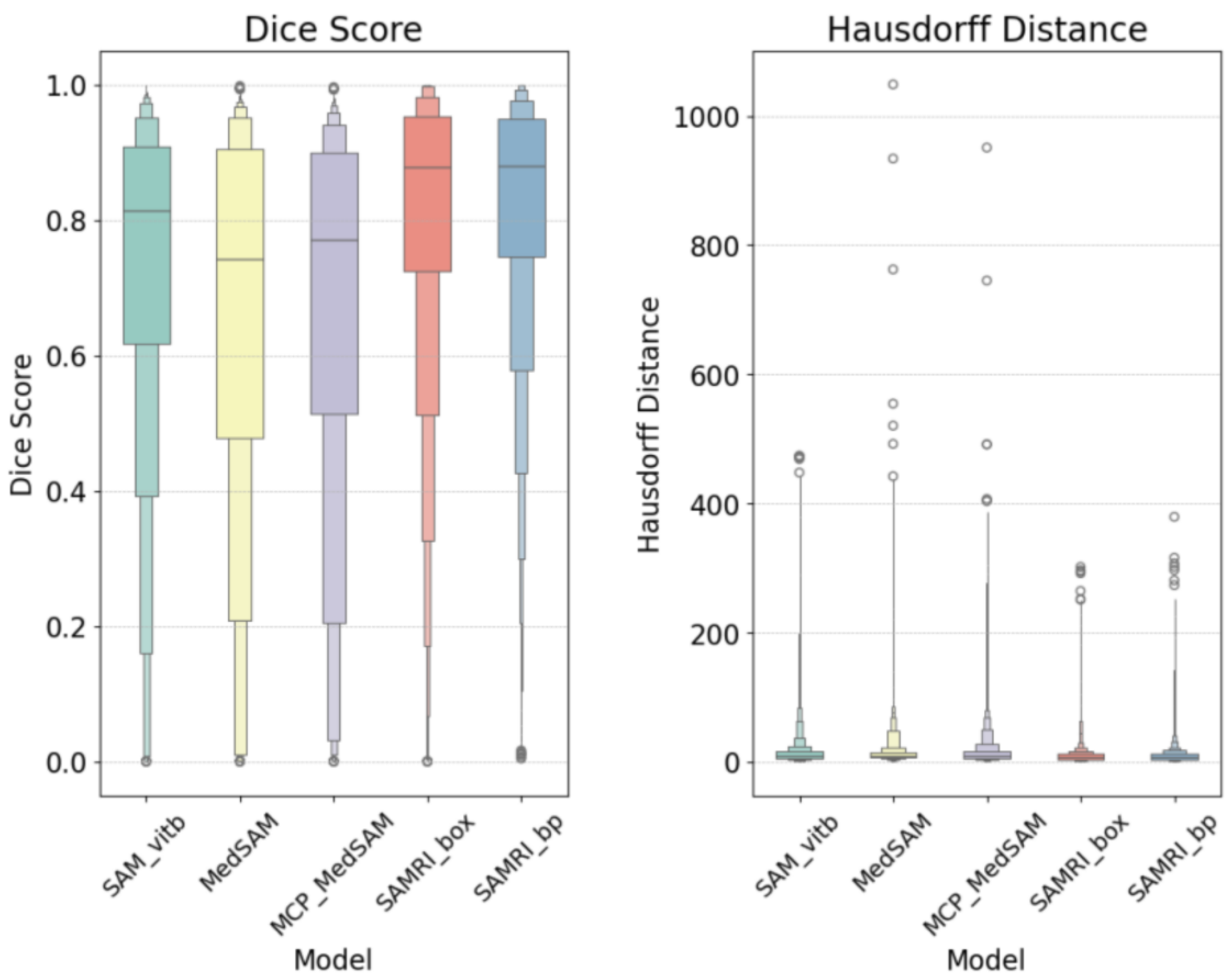

Fig. 3 Overall segmentation performance across all models. **Left:** Dice Similarity Coefficient (DSC) distributions per model across all 47 targets and 36 datasets. $\text{SAMRI}_{bp}$ achieves the highest median DSC (0.87 ± 0.11) with the tightest spread, followed closely by $\text{SAMRI}_{box}$ (0.85 ± 0.15). MCP-MedSAM (0.73 ± 0.22) and MedSAM (0.74 ± 0.24) perform similarly, both lower than SAM ViT-B in overall median DSC due to MedSAM's weaker performance on small structures. **Right:** Hausdorff Distance (HD, pixels) distributions per model. SAMRI variants achieve the lowest boundary errors, confirming superior contour precision across diverse anatomies and MRI contrasts. All pairwise differences between SAMRI variants and baselines are statistically significant (Wilcoxon signed-rank test, $p < 0.05$).

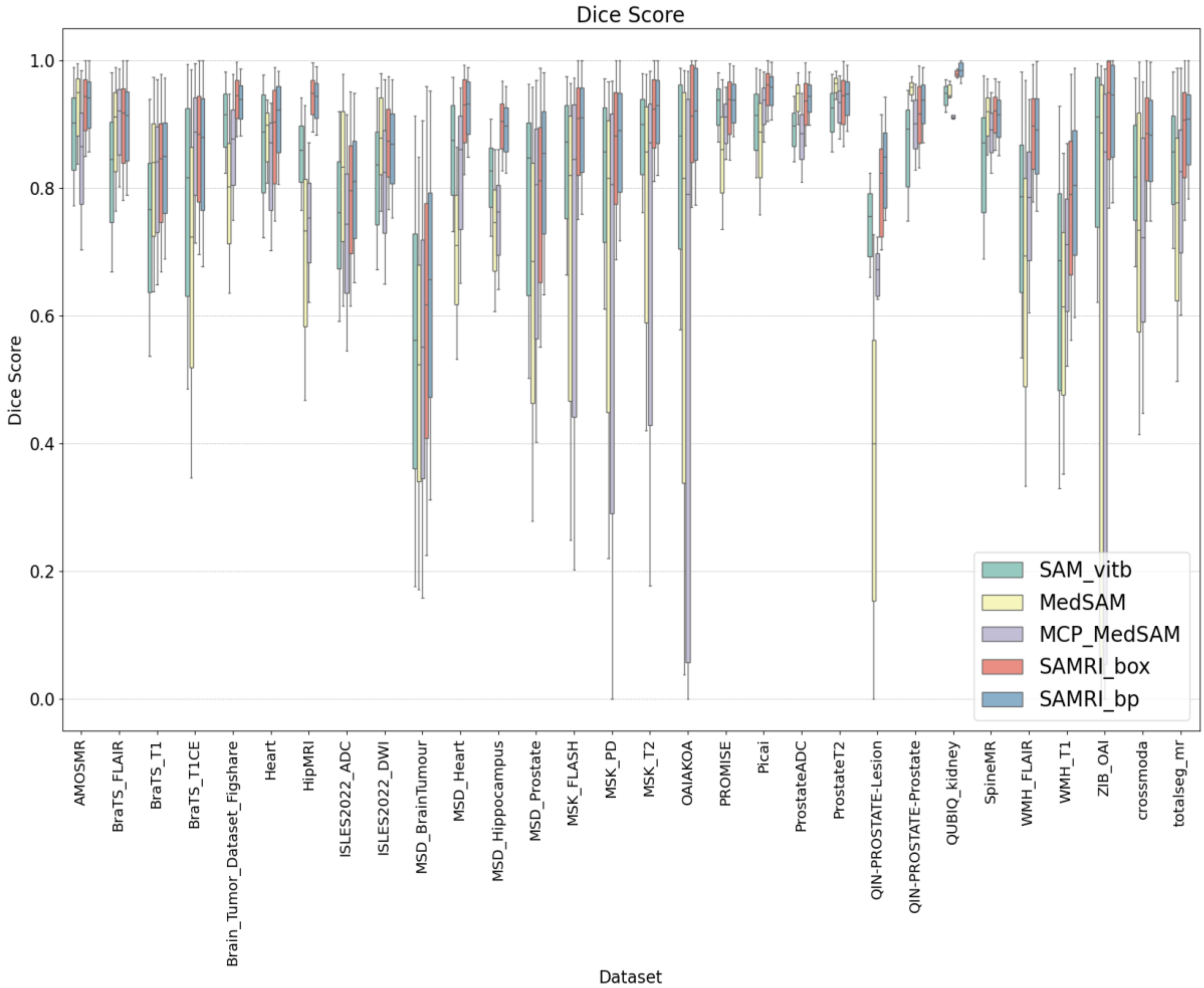

Fig. 4 Per-dataset DSC performance across all 36 datasets and five models. Each panel shows the distribution of Dice Similarity Coefficient (DSC) scores for one dataset. SAMRI variants consistently achieve higher median DSC and lower variance than baselines across the majority of datasets, with the most pronounced improvements in datasets containing small or low-contrast structures.

## Training Computational Efficiency

Fig. 5 compares the total training FLOPs required by SAMRI against representative SAM-based models, normalised to 200 epochs on 1.1 M slice–mask pairs. By precomputing image embeddings and training only the mask decoder, $\mathrm{SAMRI_{box}}$ and $\mathrm{SAMRI_{bp}}$ require only $1.21 \times 10^{20}$ FLOPs—a 99% reduction relative to MedSAM ($3.22 \times 10^{22}$). Adapter-based models (SAM-Med2D, Med-SA) achieve partial savings ($\sim 1.1 \times 10^{22}$) by freezing the encoder but still recompute features online each epoch. Distillation-based EMedSAM incurs the highest cost ($7.59 \times 10^{22}$) because the ViT-H teacher must execute a forward pass at every iteration. MCP-MedSAM, which replaces the ViT-B encoder with a lightweight TinyViT-5M backbone, requires approximately $2.58 \times 10^{21}$ FLOPs, positioning it between adapter-based and decoder-only approaches.

During training, VRAM usage can be reduced by lowering the batch size, with a practical minimum of ~2.5 GB. Precomputed image embeddings required ~2.2 TB of storage (~1.1 million image-mask pairs), and the final checkpoint is ~350 MB. Practically,

SAMRI can be trained on a single commercial GPU in around one month on million-sample datasets, indicating that the two-stage procedure enables feasible fine-tuning of a competitive model. See Table S6 in Supplementary for detailed experiments on a Mac mini.

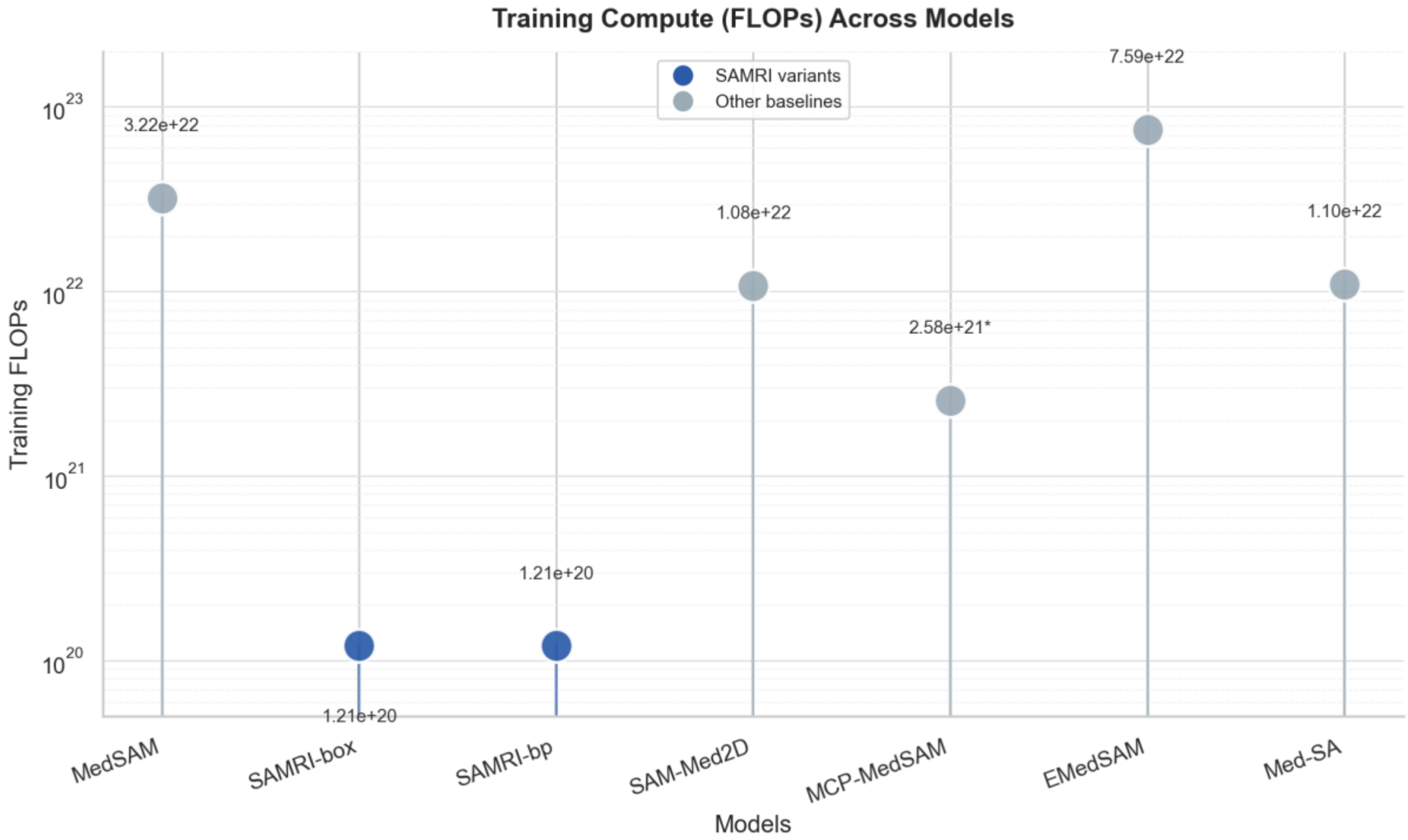

Fig. 5 Training FLOPs across models. Total training floating-point operations (FLOPs) required by each model, shown on a logarithmic scale. MedSAM (full end-to-end fine-tuning of SAM ViT-B) requires $3.22 \times 10^{22}$ FLOPs. SAM-Med2D (frozen encoder with adapters) and Med-SA require $1.08 \times 10^{22}$ and $1.10 \times 10^{22}$ FLOPs respectively, reflecting savings from partial encoder freezing. EMedSAM ($7.59 \times 10^{22}$) exceeds MedSAM due to additional distillation objectives. MCP-MedSAM uses a lightweight TinyViT-5M encoder and requires approximately $2.58 \times 10^{21}$ FLOPs. $\text{SAMRI}_{\text{box}}$ and $\text{SAMRI}_{\text{bp}}$, which train only the mask decoder on precomputed embeddings, require only $1.21 \times 10^{20}$ FLOPs—an approximate 99% reduction relative to MedSAM. FLOPs are calculated based on training for 200 epochs on 1.1 M slice–mask pairs using a Mac mini (M4) chip.

## Substantial Gains in Small-Object Segmentation

Across all size bins, SAMRI (box and box+point) outperforms SAM ViT-B, MedSAM, and MCP-MedSAM (Table 1). For small structures (<0.5% of image area), $\text{SAMRI}_{\text{bp}}$ reaches $0.84 \pm 0.26$ DSC—an absolute gain of +0.25 over MedSAM (+42.4%) and +0.23 over MCP-MedSAM ($0.61 \pm 0.44$), with $\text{SAMRI}_{\text{box}}$ similar at $0.83 \pm 0.31$. For medium-sized targets (0.5–3.5% of image area), $\text{SAMRI}_{\text{bp}}$ attains $0.85 \pm 0.22$, improving on MedSAM by +0.18 (+26.9%) and on MCP-MedSAM ($0.74 \pm 0.44$) by +0.11. Even for large objects (>3.5%), where baselines are already strong, $\text{SAMRI}_{\text{bp}}$ maintains a measurable edge ($0.95 \pm 0.12$ vs $0.92 \pm 0.16$ for MedSAM; +0.03, +3.3%); MCP-MedSAM matches MedSAM at $0.92 \pm 0.12$ for large structures but shows substantially higher variance on small targets. Overall, the largest relative gains appear in the small and medium regimes, consistent with SAMRI's design emphasis on challenging clinically important small structures

**TABLE 1** DSC performance by object size category.

| Model | Size Category* | | |
|---|---|---|---|
| | **Small (<0.5%)** | **Medium(0.5–3.5%)** | **Large(>3.5%)** |
| SAM ViT-B | $0.76 \pm 0.34$ | $0.77 \pm 0.34$ | $0.91 \pm 0.18$ |
| MedSAM | $0.59 \pm 0.47$ | $0.67 \pm 0.42$ | $0.92 \pm 0.16$ |
| MCP-MedSAM | $0.61 \pm 0.44$ | $0.74 \pm 0.44$ | $0.92 \pm 0.12$ |
| SAMRI$_{box}$ | $\mathbf{0.83 \pm 0.31}$ | $\mathbf{0.85 \pm 0.26}$ | $\mathbf{0.95 \pm 0.13}$ |
| SAMRI$_{bp}$ | $\mathbf{0.84 \pm 0.26}$ | $\mathbf{0.85 \pm 0.22}$ | $\mathbf{0.95 \pm 0.12}$ |
| $\Delta$ vs MedSAM[†] | +0.25 (+42.4%) | +0.18 (+26.9%) | +0.03 (+3.3%) |

**DSC:** Dice similarity coefficient. [*]Size bins stratified by object area (percentage of image) using the global distribution. [†]$\Delta$ and % computed for SAMRI$_{bp}$ relative to MedSAM ($\Delta$ = SAMRI$_{bp}$ − MedSAM; % = $\Delta$/MedSAM).

Fig. 6 further illustrates the relationship between segmentation performance and object size. The top left panel shows the median DSC (±IQR: Interquartile Range) across size bins. We include a dashed reference at DSC = 0.80 as a heuristic baseline. SAMRI consistently stays above this threshold and demonstrates the most significant gains for small and medium objects, where competing methods show a steep drop in accuracy.

The bottom panel presents the distribution of object sizes in the dataset, showing that the vast majority of training samples are small structures (<3.5% of image area). This highlights the practical importance of SAMRI's improvement in this regime, as small anatomical targets are both abundant in MRI data and often critical for clinical decision-making.

Notable improvements were observed in several small anatomical structures (Supplementary Table S3). For example, SAMRI achieved a DSC of 0.85 on femoral cartilage, whereas MedSAM struggled (0.00 DSC). In the extra-meatal part of a vestibular schwannoma, SAMRI reached 0.84 DSC compared to 0.57 for MedSAM, which corresponds to a 47% improvement. The left adrenal gland showed 0.85 DSC versus 0.63 (a 35% gain), and the cochlea achieved 0.86 DSC versus 0.73 (an 18% gain). These results highlight the substantial advantage of SAMRI in segmenting small, clinically significant MRI structures—domains where baseline models fail or underperform.

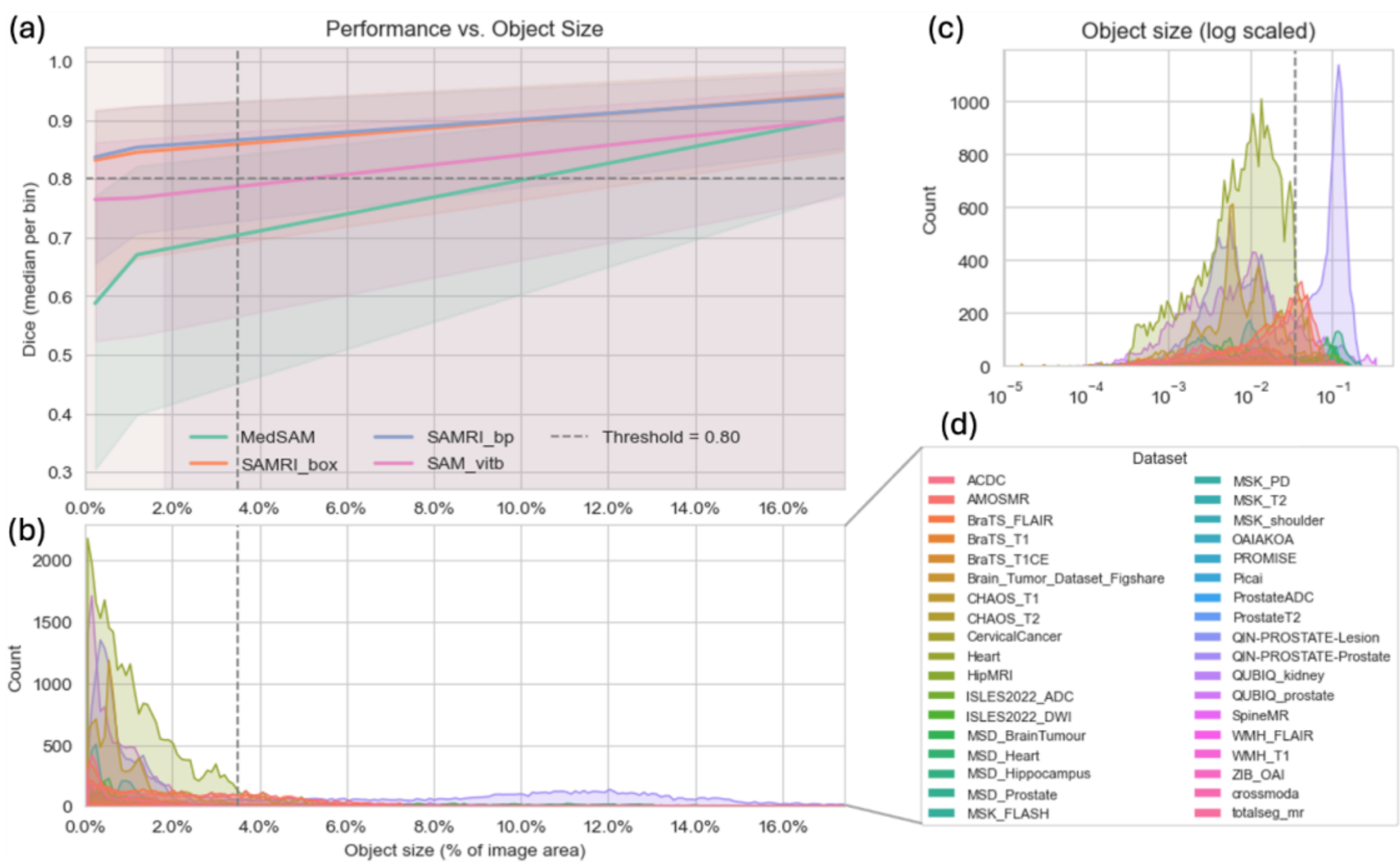

Fig. 6 Performance vs. object size. (a) Median Dice (±IQR) by object-size bin; the horizontal dashed line marks the 0.80 as a heuristic baseline. The vertical dashed line separates large objects (>3.5% of image area) from small/medium (<3.5%). SAMRI attains the highest Dice across sizes, with the largest gains in the small–medium regime. (b) Histogram of training samples by object size (same x-axis), showing a dominance of small objects. (c) Log-scaled object-size distribution for the full dataset, again indicating small-object dominance. (d) Legend for the 36 datasets.

## Qualitative Analysis

Fig. 7 presents qualitative comparisons across diverse anatomies and MRI contrasts for SAM ViT-B, MedSAM, $\text{SAMRI}_{\text{box}}$, and $\text{SAMRI}_{\text{bp}}$; MCP-MedSAM is omitted from the qualitative panel as its boundary predictions are consistent with its quantitative performance and do not introduce additional visual patterns beyond those observed for MedSAM. SAMRI consistently produces smoother and more anatomically coherent contours than MedSAM and SAM ViT-B. Specifically, baseline models exhibit two common failure modes: (1) *boundary over-extension*, where predicted contours extend beyond the true structure into adjacent tissue (e.g., Row 2 left, kidney: MedSAM's magenta contour extends into surrounding tissue; Row 3 middle, prostate: SAM ViT-B significantly over-segments the target region), and (2) *missed fine structures*, where thin or low-contrast regions are incompletely segmented (e.g., Row 1 left and middle, brain tumor: baseline models fail to capture the irregular tumor boundaries; Row 4, femur cartilage: SAM ViT-B and MedSAM miss portions of the thin cartilage structure). In contrast, SAMRI variants maintain precise boundary delineation even in these challenging regions. When structures have clear, well-defined boundaries, all models perform comparably well. For example, in Row 2 middle (left atrium) and Row 2 right (prostate), all four methods achieve similar segmentation quality, with contours closely aligned to the ground truth.

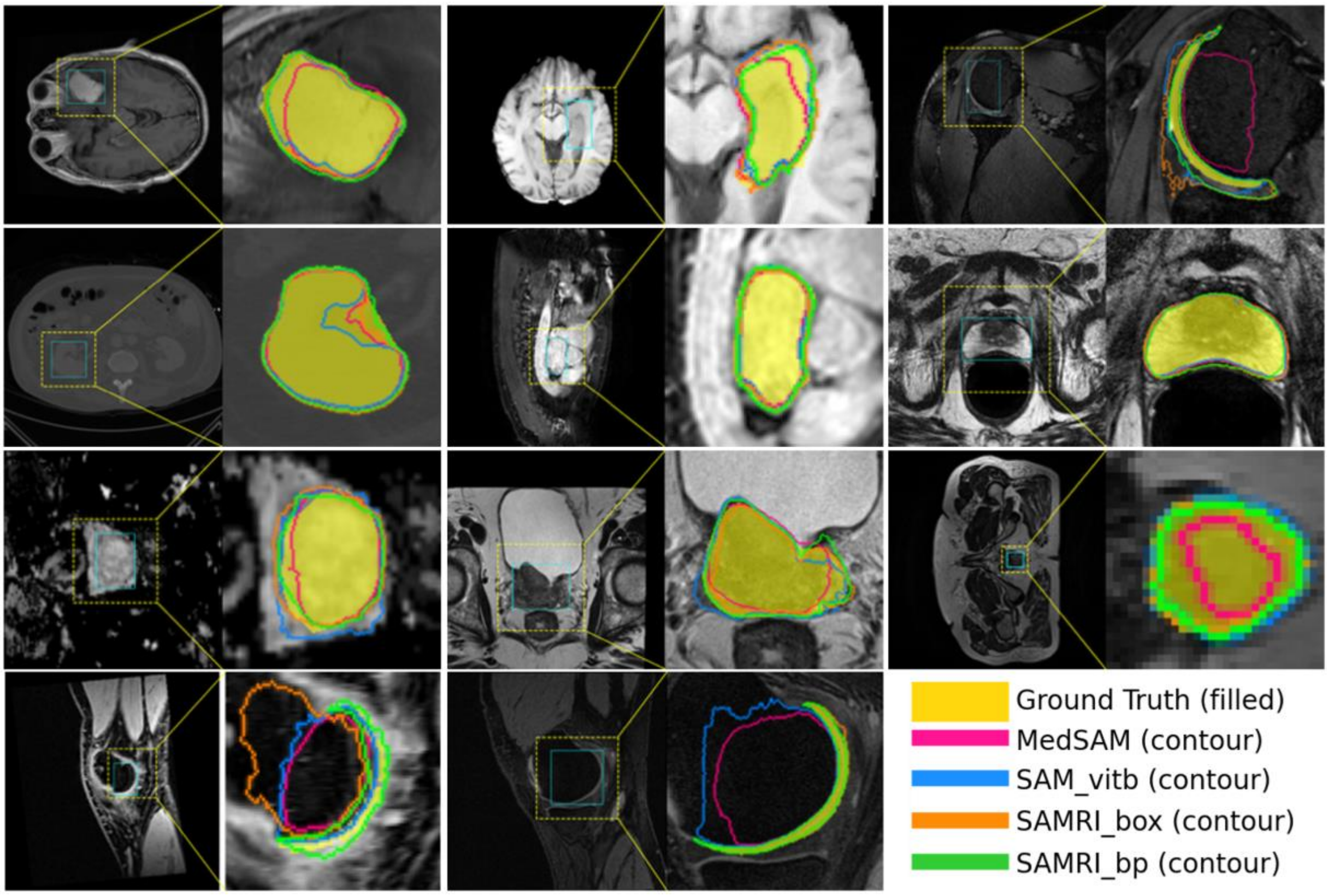

Fig 7 Qualitative comparisons across anatomies and contrasts. Each case shows the full image with a zoom-in. Ground truth (yellow fill); SAM ViT-B (blue), MedSAM (magenta), SAMRIbox (orange), SAMRIbp (green) contours. SAMRI variants adhere more closely to the reference, especially on thin or small structures and low-contrast regions; SAMRIbp most consistently captures fine extremities and concavities.

The difference between $\text{SAMRI}_{\text{box}}$ and $\text{SAMRI}_{\text{bp}}$ remains modest overall; however, the additional point prompt in $\text{SAMRI}_{\text{bp}}$ provides consistent refinement around ambiguous borders (e.g., Row 3 left and right, prostate: the green contour shows tighter adherence to ground truth than the orange contour). These visual patterns align with the quantitative results in Table 1: the most pronounced improvements occur on small and medium structures with ambiguous boundaries, while performance on large, well-defined objects remains comparable across methods.

## Zero-shot Generalization

We evaluated SAMRI's zero-shot generalization on six unseen datasets (Fig. 8). $\text{SAMRI}_{\text{bp}}$ achieved consistently strong performance across all datasets, with $\text{SAMRI}_{\text{box}}$ performing comparably. MedSAM, MCP-MedSAM, and SAM ViT-B exhibited larger variability or lower median Dice scores, with MCP-MedSAM showing notably weaker performance on small and medium objects—consistent with the dilution effect of multi-modal training on MRI-specific targets observed in the seen-set evaluation. The size-stratified view (Fig. 8 b) pinpoints the source of these gains: for small and medium targets, the SAMRI distributions shift upward with higher medians and substantially higher lower quartiles, while performance on large objects remains competitive across all models. Notably, the MSK shoulder dataset includes humerus and scapula cartilage whose properties resemble knee cartilage (see Fig. 7: upper-right: shoulder cartilage; bottom row: knee cartilage); training on knee cartilage appears to transfer effectively to these anatomically distinct yet structurally similar targets, boosting zero-shot performance. Beyond overall accuracy, SAMRI also exhibits narrower IQR and fewer catastrophic outliers, indicating improved robustness to contrast, noise, and label conventions. Overall, SAMRI lifts zero-shot accuracy without sacrificing stability, with the box+point prompt yielding the most reliable improvements.

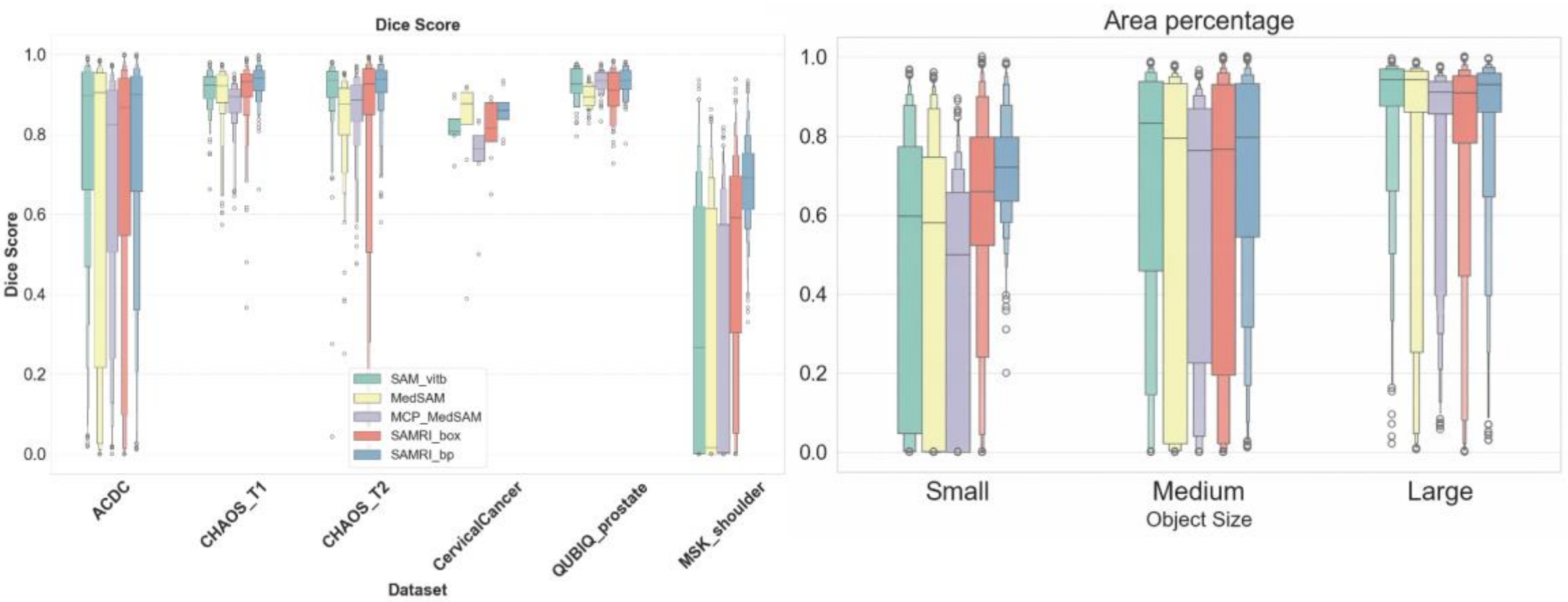

Fig. 8 Zero-shot performance across datasets and object sizes. (a) Dataset-wise Dice boxplots on six held-out (zero-shot) datasets. $SAMRI_{bp}$ achieves consistently strong performance across all datasets, with $SAMRI_{box}$ performing comparably. MedSAM, MCP-MedSAM, and SAM ViT-B exhibit larger variability and lower median Dice scores, particularly on CHAOS T2, CervicalCancer, and MSK shoulder. (b) Size-stratified boxplots (small/medium/large). SAMRI variants shift the distributions upward for small and medium targets (higher medians and lower tails) while remaining competitive on large objects. MCP-MedSAM shows substantially lower performance on small and medium structures, consistent with the dilution effect of multi-modal training on MRI-specific targets.

## Training and Validation Loss Observation

Across both box-only and box+point prompting, training loss decreases smoothly while seen-set validation loss remains largely flat, indicating no overfitting (Supplementary Fig. S14). We monitor two criteria: (i) zero-shot validation loss on a held-out 10% ACDC split and (ii) seen-set validation loss. The minimum zero-shot loss occurs at epoch 40 for both prompting regimes, whereas the seen-task minima occur later—epoch 150 for box-only and epoch 170 for box+point. To accommodate different deployment needs, we therefore retain four checkpoints: box-only @ 40 (simplest prompt; best zero-shot accuracy, i.e., lowest zero-shot loss), box-only @ 150 (simplest prompt; best seen-task accuracy), box+point @ 40 (highest zero-shot accuracy), and box+point @ 170 (highest seen-task accuracy).

## Practical deployment and interactive segmentation

We further demonstrate the practical usability of SAMRI through a user-friendly local graphical interface (Supplementary Fig. S13). The interface supports interactive visualization of volumetric MR images and prompt-based segmentation using bounding boxes and point prompts, with segmentation results displayed in real time for iterative refinement. Pretrained SAMRI models can be applied directly on standard local machines without specialized deployment infrastructure, enabling efficient and responsive interaction, typically within seconds per slice on a single commercial GPU (~4.5 GB VRAM). By abstracting preprocessing, model loading, and inference, the interface lowers the technical barrier for end users and facilitates practical adoption of SAMRI for interactive MRI segmentation.

## Discussion

This work introduces SAMRI, an MRI-specialized adaptation of SAM that updates only the mask decoder on precomputed image embeddings, enabling an efficient training pipeline. Across 47 anatomical/pathological targets and 36 datasets, SAMRI—especially the box+point variant—outperforms SAM ViT-B, MedSAM, and MCP-MedSAM in Dice and boundary metrics, with the largest gains on small and medium structures. Qualitative analyses show reduced over-segmentation into adjacent tissue and more precise boundary adherence on thin or low-contrast targets, aligning with the size-stratified improvements. SAMRI also attains strong zero-shot performance on six held-out datasets, indicating robust generalization.

**Why it works.** We attribute the gains to the following factors: (i) retaining SAM's pretrained image encoder while updating only the mask decoder allows MRI-specific adaptation without losing general visual representations; (ii) a two-stage pipeline (Fig. 1) that precomputes image embeddings and then trains the decoder avoids redundant encoder passes, substantially reducing computational cost and memory usage; (iii) training on a broad, MRI-only corpus spanning different contrasts and anatomies regularizes the decoder toward MRI-specific textures, rather than features from other modalities (e.g., CT/X-ray/ultrasound) that may not generalize to MRI. Finally, the box+point prompt enhances spatial guidance: the box provides coarse localization, while the point specifies the intended target, reducing ambiguity in small, or low-contrast regions.

**Small objects matter.** Most samples in our corpus occupy <3.5% of the image area. In that regime, baselines drop sharply in accuracy, whereas SAMRI raises the entire curve above the 0.80 threshold. These targets—cartilage, small lesions, vessels—are common and clinically important in MRI, often driving clinical decisions. SAMRI better preserves narrow tips and concavities while avoiding over-segmentation into adjacent tissue, as reflected in both quantitative margins and visual comparisons. Training on an MRI-only corpus avoids modality-specific artifacts and sharpens the decoder's inductive bias for MRI textures, explaining why SAMRI gains most where baseline priors are weakest on small, low-contrast targets. This is further corroborated by MCP-MedSAM [47], which introduces learnable modality and content prompts but trains across 11 imaging modalities: it achieves a mean DSC of $0.73 \pm 0.22$—with substantially higher variance than $\text{SAMRI}_{\text{bp}}$ ($\pm 0.11$)—and exhibits near-zero performance on musculoskeletal cartilage (femoral DSC: 0.02; scapular DSC: 0.00), reinforcing that multi-modal training dilutes MRI-specific fidelity for small, structurally demanding targets.

**Efficiency and practicality.** Training completes in ~76 hours on 8× MI300X (~608 GPU-hours), and inference requires only ~4.5 GB of VRAM. This represents a substantial improvement over full-model retraining and adapter-based methods that recompute image features every epoch, making SAMRI practical for typical academic and clinical compute environments.

**Zero-shot generalization.** On unseen datasets, SAMRI maintains higher median Dice scores and smaller variance than comparators. Gains are largest for small and medium objects, while performance on large structures remains competitive, evidencing that decoder specialization enhances adaptation to MRI statistics without compromising generalization across scanners, contrasts, or labeling conventions. This robustness is crucial for deployment in heterogeneous clinical settings where retraining is costly or impractical. Notably, training solely on knee cartilage substantially improves zero-shot performance on shoulder cartilage, indicating that exposure to diverse yet structurally related examples enhances transfer. More broadly, these results support our choice to train on a large, MRI-only corpus spanning sequences and anatomies, rather than narrow, modality-specific subsets (as in many adapter-based baselines in Supplementary Table S2), to maximize zero-shot generalization.

**Limitations.** First, our pipeline supports 2D slices currently, foregoing inter-plane context that can help disambiguate boundaries. Second, despite broad coverage, dataset composition may still under-represent certain vendors, rare pathologies, or pediatric populations; residual domain shift is possible. Third, label noise and inter-rater variability—ubiquitous in public MRI sets—can cap achievable metrics and may differentially affect small-object evaluations (Supplementary Fig. S12). Fourth, while embedding precomputation accelerates training, it requires ~2.2 TB of storage at full scale; clinics with constrained storage may need cloud or streamed variants. Finally, we did not study uncertainty calibration or failure detection, both critical for safe clinical use.

**Future work.** Extending SAMRI to 2.5D or 3D architectures with explicit inter-slice context may further improve robustness, particularly for thin structures and regions with ambiguous boundaries. On the data side, we plan to expand pretraining and fine-tuning to broader, multi-institutional MRI cohorts to enhance out-of-domain generalization. While a graphical user interface is already provided, the current implementation is based on a Jupyter notebook environment and may offer lower runtime efficiency than standalone desktop applications; future work will focus on developing a system-level and web-server-based interface to further improve usability and deployment flexibility.

## Conclusion

SAMRI is an MRI-specialized adaptation of SAM that fine-tunes only the mask decoder on precomputed embeddings, trained on a large, multi-sequence MRI resource spanning whole-body organs and clinically relevant pathologies. It achieves competitive accuracy, especially for small, clinically salient structures, while remaining computationally frugal and deployment-ready. The combination of strong zero-shot generalization, training efficiency, and practical VRAM requirements makes SAMRI a viable foundation for MRI segmentation in real-world settings. Importantly, SAMRI could materially aid clinical researchers by enabling semi-automatic annotation, streamlining workflows that require manual corrections, and improving robustness for models intended for clinical use—especially in data-scarce settings such as rare diseases.

## Ethics statement

All datasets used in this study are publicly available and were originally collected with appropriate ethical approvals by the respective dataset providers. No new human or animal data were collected for this study.

## Declaration of generative AI and AI-assisted technologies

During the preparation of this work the author(s) used ChatGPT in order to check grammar and refine language. After using this tool/service, the author(s) reviewed and edited the content as needed and take(s) full responsibility for the content of the published article.

## Data availability

All datasets used in this study are publicly available from their respective sources, as detailed in Supplementary Table S1. Code and pretrained models are publicly available at https://github.com/wangzhaomxy/SAMRI.

## Author contributions

Zhao Wang conceived the study, developed the methodology, curated the data, conducted the formal analysis and investigation, developed the software, provided resources, administered the project, created the visualizations, and wrote the original manuscript. Wei Dai contributed to conceptualization, data curation, formal analysis, investigation, methodology development, and manuscript review and editing. Thuy Thanh Dao and Steffen Bollmann contributed to software development, validation, and manuscript review and editing. Hongfu Sun contributed to methodology development, supervision, and manuscript review and editing. Craig Engstrom contributed to data curation, funding acquisition, resource provision, supervision, and manuscript review and editing. Shekhar S. Chandra conceived the study, developed the methodology, curated the data, acquired funding, provided resources, administered the project, supervised the research, and contributed to manuscript review and editing.

## Supporting Information

**Figure S1.** Proportion of image–mask pairs originating from new datasets versus those inherited from MedSAM's collection, illustrating the expanded anatomical and contrast diversity introduced in SAMRI's training corpus.

**Figures S2–S8.** Per-target and per-dataset DSC results for all 47 segmentation targets and 36 datasets, showing detailed model comparisons across anatomies and MRI contrasts.

**Figure S9.** Comparison of SAM ViT-H and SAM ViT-B on our development set, showing no consistent advantage for ViT-H in zero-shot MRI segmentation.

**Figure S10.** Generalization evaluation of adapter-based baselines (Med-SA, SAM-Med2D, SAMed) using publicly released checkpoints across our full range of MRI contrasts, anatomies, and tasks.

**Figure S11.** Effect of discarding training masks smaller than 10% of the largest slice's mask area, illustrating improved training stability from this filtering step.

**Figure S12.** Analysis of label noise and inter-rater variability in public MRI datasets and its potential impact on small-object evaluation metrics.

**Figure S13.** Screenshot and walkthrough of the SAMRI graphical user interface, demonstrating interactive volumetric MRI visualization and prompt-based segmentation with bounding-box and point prompts.

**Figure S14.** Loss trajectories across epochs for box-only and box+point prompting regimes, showing smooth convergence of training loss and stable validation loss without overfitting.

**Table S1.** Detailed breakdown of all 36 MRI datasets used in this study, including dataset name, source, segmentation targets, MRI contrast, number of subjects, and train/validation/test/zero-shot assignment.

**Table S2.** Overview of existing SAM adaptation strategies for medical image segmentation, covering method category, key design choices, training modalities, and reported performance.

**Tables S3–S4.** Full per-target (Table S3) and per-dataset (Table S4) quantitative results (DSC, HD, MSD) for all evaluated models.

**Table S5.** Computational cost comparison of SAM's image encoder versus mask decoder, motivating the decoder-only training strategy.

**Table S6.** Training and inference experiments on a Mac mini (M4 chip), reporting wall-clock time, VRAM usage, and FLOPs across model configurations.

**Table S7.** Training resource requirements across models, standardised by GPU, dataset size, and training epochs, reporting total GPU hours.

**Appendix S1.** Derivation of training efficiency gains for the two-stage pipeline, with formula-level analysis of the reduction in wall-clock training time achieved by precomputing image embeddings.

**Appendix S2.** Definitions of the focal loss ($\mathcal{L}_{\text{Focal}}$) and Dice loss ($\mathcal{L}_{\text{Dice}}$) components used in the combined SAMRI training objective.

**Appendix S3.** Definition and computation of evaluation metrics: Dice Similarity Coefficient (DSC), Hausdorff Distance (HD), and Mean Surface Distance (MSD).

# References

**1.** Fernandes SL, Tanik UJ, Rajinikanth V, Karthik KA. A reliable framework for accurate brain image examination and treatment planning based on early diagnosis support for clinicians. Neural Comput Appl. 2020;32(20):15897–15908. https://doi.org/10.1007/s00521-019-04369-5.

**2.** Joskowicz L, Cohen D, Caplan N, Sosna J. Inter-observer variability of manual contour delineation of structures in CT. Eur Radiol. 2019;29(3):1391–1399. https://doi.org/10.1007/s00330-018-5695-5.

**3.** Filippi M, Agosta F. Magnetization transfer MRI in multiple sclerosis. J Neuroimaging. 2007;17:22S-26S.

**4.** Jager F, Hornegger J. Nonrigid Registration of Joint Histograms for Intensity Standardization in Magnetic Resonance Imaging. IEEE Trans Med Imaging. 2009;28(1):137–150. https://doi.org/10.1109/TMI.2008.2004429.

**5.** Trattnig S, Hangel G, Robinson SD, Juras V, Szomolanyi P, Dal-Bianco A. Ultrahigh-field MRI: where it really makes a difference. Die Radiologie. 2024;64(1):1–8. https://doi.org/10.1007/s00117-023-01184-x.

**6.** Springer E, Dymerska B, Cardoso PL, et al. Comparison of Routine Brain Imaging at 3 T and 7 T. Invest Radiol. 2016;51(8). https://journals.lww.com/investigativeradiology/fulltext/2016/08000/comparison_of_routine_brain_imaging_at_3_t_and_7_t.1.aspx.

**7.** Lee HH, Gu Y, Zhao T, et al. Foundation models for biomedical image segmentation: A survey. arXiv preprint arXiv:2401.07654; 2024.

**8.** Kraff O, Fischer A, Nagel AM, Mönninghoff C, Ladd ME. MRI at 7 Tesla and above: demonstrated and potential capabilities. J Magn Reson Imaging. 2015;41(1):13–33.

**9.** Harada T, Kudo K, Fujima N, et al. Quantitative susceptibility mapping: basic methods and clinical applications. Radiographics. 2022;42(4):1161–1176.

**10.** Haacke EM, Chen Y, Utriainen D, et al. STrategically Acquired Gradient Echo (STAGE) imaging, part III: Technical advances and clinical applications of a rapid multi-contrast multi-parametric brain imaging method. Magn Reson Imaging. 2020;65:15–26. https://doi.org/10.1016/j.mri.2019.09.006.

**11.** Finnis KW, Starreveld YP, Parrent AG, Sadikot AF, Peters TM. Application of a Population Based Electrophysiological Database to the Planning and Guidance of Deep Brain Stereotactic Neurosurgery. In: Dohi T, Kikinis R, eds. Medical Image Computing and Computer-Assisted Intervention — MICCAI 2002. Springer Berlin Heidelberg; 2002:69–76.

**12.** Nouranian S, Mahdavi SS, Spadinger I, Morris WJ, Salcudean SE, Abolmaesumi P. A Multi-Atlas-Based Segmentation Framework for Prostate Brachytherapy. IEEE Trans Med Imaging. 2015;34(4):950–961. https://doi.org/10.1109/TMI.2014.2371823.

**13.** Dowling JA, Sun J, Pichler P, et al. Automatic Substitute Computed Tomography Generation and Contouring for Magnetic Resonance Imaging (MRI)-Alone External Beam Radiation Therapy From Standard MRI Sequences. Int J Radiat Oncol Biol Phys. 2015;93(5):1144–1153. https://doi.org/10.1016/j.ijrobp.2015.08.045.

**14.** Chandra SS, Dowling JA, Shen KK, et al. Patient Specific Prostate Segmentation in 3-D Magnetic Resonance Images. IEEE Trans Med Imaging. 2012;31(10):1955–1964. https://doi.org/10.1109/TMI.2012.2211377.

**15.** Chua LO. CNN: A Vision of Complexity. Int J Bifurcat Chaos. 1997;07(10):2219–2425. https://doi.org/10.1142/s0218127497001618.

**16.** Azad R, Aghdam EK, Rauland A, et al. Medical Image Segmentation Review: The Success of U-Net. IEEE Trans Pattern Anal Mach Intell. 2024;46(12):10076–10095. https://doi.org/10.1109/TPAMI.2024.3435571.

**17.** Çiçek O, Abdulkadir A, Lienkamp SS, Brox T, Ronneberger O. 3D U-Net: learning dense volumetric segmentation from sparse annotation. In: International Conference on Medical Image Computing and Computer-Assisted Intervention. Springer; 2016:424–432.

**18.** Zhu Y, Tchetchenian A, Yin X, Liew A, Song Y, Meijering E. AC-UNet: Adaptive Connection UNet for White Matter Tract Segmentation Through Neural Architecture Search. In: 2024 IEEE International Symposium on Biomedical Imaging (ISBI); 2024:1–5. https://doi.org/10.1109/ISBI56570.2024.10635404.

**19.** Harirpoush A, Rasoulian A, Kersten-Oertel M, Xiao Y. Architecture Analysis and Benchmarking of 3D U-Shaped Deep Learning Models for Thoracic Anatomical Segmentation. IEEE Access. 2024;12:127592–127603. https://doi.org/10.1109/ACCESS.2024.3456674.

**20.** Oktay O, Schlemper J, Folgoc LL, et al. Attention u-net: Learning where to look for the pancreas. arXiv preprint arXiv:1804.03999; 2018.

**21.** Dai W, Woo B, Liu S, et al. CAN3D: Fast 3D medical image segmentation via compact context aggregation. Med Image Anal. 2022;82:102562.

**22.** Zhang J, Zhang Y, Jin Y, Xu J, Xu X. Mdu-net: Multi-scale densely connected u-net for biomedical image segmentation. Health Inf Sci Syst. 2023;11(1):13.

**23.** Isensee F, Petersen J, Klein A, et al. nnu-net: Self-adapting framework for u-net-based medical image segmentation. arXiv preprint arXiv:1809.10486; 2018.

**24.** Alom MZ, Yakopcic C, Hasan M, Taha TM, Asari VK. Recurrent residual U-Net for medical image segmentation. J Med Imaging. 2019;6(1):014006–014006.

**25.** Milletari F, Navab N, Ahmadi SA. V-net: Fully convolutional neural networks for volumetric medical image segmentation. In: 2016 Fourth International Conference on 3D Vision (3DV). IEEE; 2016:565–571.

**26.** Ronneberger O, Fischer P, Brox T. U-Net: Convolutional Networks for Biomedical Image Segmentation. In: Medical Image Computing and Computer-Assisted Intervention – MICCAI 2015. Springer International Publishing; 2015:234–241.

**27.** Stringer C, Wang T, Michaelos M, Pachitariu M. Cellpose: a generalist algorithm for cellular segmentation. Nat Methods. 2021;18(1):100–106. https://doi.org/10.1038/s41592-020-01018-x.

**28.** Kushol R, Parnianpour P, Wilman AH, Kalra S, Yang YH. Effects of MRI scanner manufacturers in classification tasks with deep learning models. Sci Rep. 2023;13(1):16791. https://doi.org/10.1038/s41598-023-43715-5.

**29.** Nomura Y, Hanaoka S, Hayashi N, et al. Performance changes due to differences among annotating radiologists for training data in computerized lesion detection. Int J Comput Assist Radiol Surg. 2024;19(8):1527–1536. https://doi.org/10.1007/s11548-024-03136-9.

**30.** Renard F, Guedria S, Palma ND, Vuillerme N. Variability and reproducibility in deep learning for medical image segmentation. Sci Rep. 2020;10(1):13724. https://doi.org/10.1038/s41598-020-69920-0.

**31.** Ren W, Tang Y, Sun Q, Zhao C, Han QL. Visual Semantic Segmentation Based on Few/Zero-Shot Learning: An Overview. IEEE/CAA J Autom Sin. 2024;11(5):1106–1126. https://doi.org/10.1109/JAS.2023.123207.

**32.** Salehi AW, Khan S, Gupta G, et al. A Study of CNN and Transfer Learning in Medical Imaging: Advantages, Challenges, Future Scope. Sustainability. 2023;15(7):5930. https://doi.org/10.3390/su15075930.

**33.** Yao W, Bai J, Liao W, Chen Y, Liu M, Xie Y. From CNN to Transformer: A Review of Medical Image Segmentation Models. J Imaging Inform Med. 2024;37(4):1529–1547. https://doi.org/10.1007/s10278-024-00981-7.

**34.** Kirillov A, Mintun E, Ravi N, et al. Segment anything. In: Proceedings of the IEEE/CVF International Conference on Computer Vision; 2023:4015–4026.

**35.** Pont-Tuset J, Van Gool L. Boosting object proposals: From pascal to coco. In: Proceedings of the IEEE International Conference on Computer Vision; 2015:1546–1554.

**36.** Li W, Hsu CY, Wang S, et al. Segment Anything Model Can Not Segment Anything: Assessing AI Foundation Model's Generalizability in Permafrost Mapping. Remote Sens. 2024;16(5):797. https://www.mdpi.com/2072-4292/16/5/797.

**37.** Mazurowski MA, Dong H, Gu H, Yang J, Konz N, Zhang Y. Segment anything model for medical image analysis: an experimental study. Med Image Anal. 2023;89:102918.

**38.** Huang Y, Yang X, Liu L, et al. Segment anything model for medical images?. Med Image Anal. 2024;92:103061. https://doi.org/10.1016/j.media.2023.103061.

**39.** Sack I. Magnetic resonance elastography from fundamental soft-tissue mechanics to diagnostic imaging. Nat Rev Phys. 2023;5(1):25–42. https://doi.org/10.1038/s42254-022-00543-2.

**40.** Ma J, He Y, Li F, Han L, You C, Wang B. Segment anything in medical images. Nat Commun. 2024;15(1):654. https://doi.org/10.1038/s41467-024-44824-z.

**41.** Nouman M, Khoriba G, Rashed EA. Rethinking MedSAM: Performance Discrepancies in Clinical Applications. 2024 IEEE International Conference on Future Machine Learning and Data Science (FMLDS). 2024:301–307. https://doi.org/10.1109/FMLDS63805.2024.00061.

**42.** Dong G, Wang Z, Chen Y, et al. An efficient segment anything model for the segmentation of medical images. Sci Rep. 2024;14(1):19425.

**43.** Chen C, Miao J, Wu D, et al. Ma-sam: Modality-agnostic sam adaptation for 3d medical image segmentation. Med Image Anal. 2024;98:103310.

**44.** Wu J, Wang Z, Hong M, et al. Medical SAM adapter: Adapting segment anything model for medical image segmentation. Med Image Anal. 2025;102:103547. https://doi.org/10.1016/j.media.2025.103547.

**45.** Gao Y, Xia W, Hu D, Wang W, Gao X. Desam: Decoupled segment anything model for generalizable medical image segmentation. In: International Conference on Medical Image Computing and Computer-Assisted Intervention. Springer; 2024:509–519.

**46.** Lei W, Xu W, Li K, Zhang X, Zhang S. MedLSAM: Localize and segment anything model for 3D CT images. Med Image Anal. 2025;99:103370.

**47.** Lyu D, Gao R, Staring M. MCP-MedSAM: A Powerful Lightweight Medical Segment Anything Model Trained with a Single GPU in Just One Day. Machine Learning for Biomedical Imaging. 2025;3:135–151. https://doi.org/10.59275/j.melba.2025-4849.

**48.** Lin TY, Goyal P, Girshick R, He K, Dollár P. Focal loss for dense object detection. In: Proceedings of the IEEE International Conference on Computer Vision; 2017:2980–2988.

**49.** Sudre CH, Li W, Vercauteren T, Ourselin S, Jorge Cardoso M. Generalised dice overlap as a deep learning loss function for highly unbalanced segmentations. In: Deep Learning in Medical Image Analysis and Multimodal Learning for Clinical Decision Support. Springer; 2017:240–248.

# Supplementary

**Table S1: 36 MRI datasets**

| Dataset Name | Modality | Segmentation Targets | # of samples | Labeled_Slices | Num_of_pairs |
|---|---|---|---|---|---|
| ACDC *# | MR | Heart anatomies | 300 | 2839 | 7991 |
| AMOS_MR # | MR | 15 abdominal organs | 60 | 5063 | 28806 |
| Brain Tumor Dataset Figshare # | MR-T1ce | Brain tumor | 3064 | 4120 | 4120 |
| BraTS FLAIR # | MR-FLAIR | Brain tumor | 954 | 60407 | 60407 |
| BraTS T1 # | MR-T1 | Brain tumor | 954 | 60407 | 60407 |
| BraTS T1CE # | MR-T1ce | Brain tumor | 899 | 26782 | 26782 |
| CC-Tumor-Heterogeneity *# | MR | Cervical Cancer | 7 | 56 | 56 |
| CHAOS T1 *# | MR-T1 | Liver, kidney, spleen | 20 | 443 | 1049 |
| CHAOS T2 *# | MR-T2 | Liver, kidney, spleen | 20 | 457 | 1068 |
| crossMoDA # | MR | Brain tumor | 227 | 2375 | 4759 |
| heart-HVSMR | MR | Heart | 20 | 1332 | 1332 |
| HipMRI | MR | Hip | 211 | 26719 | 108216 |
| ISLES2022-ADC # | MR-ADC | Ischemic stroke lesion | 235 | 3044 | 3444 |
| ISLES2022-DWI # | MR-DWI | Ischemic stroke lesion | 235 | 3044 | 3444 |
| MSD-Heart # | MR | Left atrial | 20 | 1330 | 1330 |
| MSD-Prostate # | MR | Prostate | 32 | 944 | 1494 |
| MSD-BrainTumour | MR-FLAIR, MR-T1w, MR-t1gd, MR-T2w | Brain tumor | 484 | 129606 | 283150 |
| MSD-Hippocampus | MR | Hippocampus | 260 | 5935 | 7619 |
| MSK knee_FLASH † | MR-FLASH | Knee | 61 | 4902 | 20834 |
| MSK knee_PD † | MR-PD | Knee | 23 | 1735 | 7351 |
| MSK knee_T2 † | MR-T2 | Knee | 63 | 5044 | 21509 |
| MSK shoulder †* | MR-T2 | Shoulder | 24 | 3645 | 7017 |
| OAI_AKOA | MR-T2 | Knee | 38 | 4466 | 20458 |
| ProstateADC | MR-ADC | Prostate | 285 | 3917 | 3917 |
| ProstateT2 | MR-T2 | Prostate | 338 | 4951 | 4951 |
| PROMISE # | MR-T2 | Prostate | 80 | 1196 | 1196 |
| Picai # | MR-bp | Prostate cancer | 1315 | 19842 | 19842 |
| QIN-PROSTATE-Lesion # | MR | Lesion | 65 | 187 | 187 |
| QIN-PROSTATE-Prostate # | MR | Prostate | 90 | 1114 | 1114 |
| QUBIQ-kidney | MR | kidney | 20 | 72 | 72 |
| QUBIQ-prostate *# | MR | Prostate | 48 | 649 | 649 |
| SpineMR # | MR | Vertebrae | 172 | 2169 | 29403 |
| TotalSeg-MR | MR | All body | 263 | 45465 | 149977 |
| WMH-FLAIR # | MR-FLAIR | White matter hyper-intensities | 170 | 3223 | 5702 |
| WMH-T1 # | MR-T1 | White matter hyper-intensities | 170 | 3223 | 5702 |
| ZIB OAI | MR-DESS | Knee | 507 | 125036 | 221447 |
| Total | | | 11734 | 565739 | 1126802 |

**Dataset reference:** ACDC [51]; AMOS_MR [52]; Brain Tumor Dataset Figshare [53–54]; BraTS [55–59]; CC-Tumor-Heterogeneity [60]; CHAOS [61]; crossMoDA [62]; heart-HVSMR [63]; ISLES2022_ADC [64]; MSD [65–66]; OAI [67]; Prostate [68]; PROMISE [69]; Picai [70]; QIN-PROSTATE [65,71]; QUBIQ [72]; SpineMR [73]; TotalSeg-MR [74]; WMH [75]

**Note:** * zero shot datasets. # mark Datasets also used by MedSAM. † private datasets.

**Table S2: Model comparison overview.**

| Model | Training scope | Methods | MRI data | Limitations |
|---|---|---|---|---|
| MedSAM | 1.5M medical images across modalities, including MRI, CT, endoscopy, ultrasound, X-ray, pathology, fundus, etc. | Fine-tune image encoder and mask decoder | Yes | High training cost; inconsistent MRI results. |
| Med-SA | BTCV (CT), REFUGE2 (optic RGB), BraTS2021 (MRI for brain), TNMIX (ultrasound), ISIC2019 (skin RGB). | Adapter encoder and decoder | Yes | Only validated on brain glioblastoma for MRI. |
| EMedSAM | FLARE2022 (CT), lung CT, X-ray, BraTS (MRI for brain). | Distillation on image encoder | Yes | Only validated on brain glioblastoma for MRI. |
| DeSAM | CHAOS (CT+MRI), NCI-ISBI (prostate MRI), I2CVB (prostate cancer MRI), PROMISE12 (prostate MRI). | Prompt enhanced on original SAM | Yes | Uses original SAM weights. |
| MedLSAM | 16 CT datasets: GLIA, ACRIN, QPC-Radiomics, Head-Neck-PET-CT, HNSCC, autoPET, MELA, LIDC-IDRI, STOIC2021, MSD-Lung, CBIS-DDSM, AMOS2022, Kits19, MSD-Colon, MSD-Pancreas, FLARE2022. | Prompt enhanced on original SAM | No | Uses original SAM weights. |
| MCP-MedSAM | Large-scale multi-modal dataset spanning 11 imaging modalities (MRI, CT, X-ray, ultrasound, etc.), trained on a single GPU within one day. | Replaces SAM's prompt encoder with learnable modality embeddings and content prompts uses lightweight TinyViT-5M image encoder. | Yes | Multi-modal training dilutes MRI-specific fidelity; substantially lower performance on small structures |
| SAM-Med2D | REFUGE2 (optic RGB), BraTS (MRI for brain), HAM10000 (skin RGB), SARTAJ, Br35H (MRI for brain), COVID19 radiography, LC25000. | Adapter encoder and decoder | Yes | MRI evaluation limited to brain tumor datasets. |
| MA-SAM | BTCV (CT), NCI-ISBI (prostate MRI), I2CVB (prostate cancer MRI), PROMISE12 (prostate MRI), EndoVis18 (robotic scene RGB), MSD-Pancreas (CT), MSD-Colon (CT), AMOS22 (CT+MRI). | Adapter for encoder; trained encoder and decoder; 3D | Yes | MRI evaluation limited to prostate datasets. |
| **SAMRI (Ours)** | **36 MRI datasets covering 47 tasks and 1.1M image–mask pairs spanning whole-body organs and clinically relevant pathologies.** | **Fine-tune mask decoder only** | **Yes** | **–** |

References: MedSAM [40]; EMedSAM [42]; MA-SAM [43]; Med-SA [44]; DeSAM [45]; MedLSAM [46]; SAM-Med2D [76]; MCP-MedSAM [47].
Note: "MRI data" indicates whether MRI data were included in training or evaluation.

**Table S3 Quantitative performance of SAMRI compared with SAM vit-b and MedSAM across 47 anatomical and pathological targets (DSC, HD, and MSD)**

| Target | Dice↑ | | | | HD↓ | | | | MSD↓ | | | |
|---|---|---|---|---|---|---|---|---|---|---|---|---|
| | SAM-ViTB | MedSAM | SAMRI$_{box}$ | SAMRI$_{bp}$ | SAM- ViTB | MedSAM | SAMRI$_{box}$ | SAMRI$_{bp}$ | SAM- ViTB | MedSAM | SAMRI$_{box}$ | SAMRI$_{bp}$ |
| Anterior Hippocampus | 0.83 (0.77, 0.86)*† | 0.77 (0.70, 0.81)*† | **0.91 (0.87, 0.94)** | 0.90 (0.86, 0.93) | 3.16 (2.83, 4.12)*† | 3.61 (2.83, 4.24)*† | **1.94 (1.37, 2.17)** | 2.00 (2.00, 2.83) | 0.94 (0.73, 1.10)*† | 1.15 (0.95, 1.37)*† | **0.56 (0.44, 0.72)** | 0.62 (0.49, 0.77) |
| Aorta | 0.94 (0.92, 0.96)*† | 0.97 (0.95, 0.98)*† | **0.98 (0.96, 0.99)** | 0.97 (0.96, 0.98) | 1.41 (1.00, 2.00)*† | 1.00 (1.00, 1.00)† | **0.97 (0.97, 1.37)** | 1.00 (1.00, 1.41) | 0.50 (0.37, 0.69)*† | **0.28 (0.20, 0.38)*†** | 0.30 (0.22, 0.48) | 0.30 (0.22, 0.50) |
| Bladder | 0.92 (0.89, 0.94)*† | 0.83 (0.71, 0.90)*† | **0.97 (0.94, 0.98)** | 0.96 (0.94, 0.98) | 2.00 (1.41, 2.83)*† | 4.00 (2.24, 5.10)*† | **1.37 (0.97, 2.17)** | 1.41 (1.00, 2.24) | 0.71 (0.54, 0.98)*† | 1.38 (0.92, 2.17)*† | 0.47 (0.29, 0.85) | **0.47 (0.31, 0.80)** |
| Bone | 0.86 (0.81, 0.89)*† | 0.73 (0.58, 0.81)*† | **0.95 (0.91, 0.97)** | 0.94 (0.91, 0.96) | 3.61 (2.83, 5.10)*† | 6.32 (4.47, 8.06)*† | **2.17 (1.37, 3.88)** | 2.24 (1.41, 4.00) | 1.45 (1.01, 1.89)*† | 2.45 (1.79, 3.41)*† | **0.75 (0.52, 1.04)** | 0.78 (0.55, 1.09) |
| Bone(Femur) | 0.96 (0.93, 0.98)*† | 0.95 (0.93, 0.97)*† | **1.00 (0.98, 1.00)** | 0.99 (0.97, 1.00) | 13.00 (6.08, 23.54)*† | 12.53 (8.06, 18.03)*† | **7.39 (4.85, 11.07)** | 8.00 (5.10, 12.04) | 2.37 (1.18, 3.86)*† | 2.84 (1.96, 3.90)*† | **1.34 (1.02, 1.92)** | 1.45 (1.09, 2.03) |
| Bone(Tibia) | 0.97 (0.95, 0.98)*† | 0.96 (0.94, 0.97)*† | **1.00 (0.97, 1.00)** | 0.99 (0.97, 1.00) | 6.32 (4.12, 10.20)*† | 7.81 (5.83, 11.00)*† | **5.23 (3.88, 7.83)** | 5.83 (4.12, 8.49) | 1.33 (0.97, 1.85)*† | 1.80 (1.40, 2.38)*† | **1.08 (0.84, 1.53)** | 1.16 (0.89, 1.61) |
| Brain Tumor | 0.81 (0.69, 0.89)*† | 0.86 (0.73, 0.92)*† | **0.89 (0.79, 0.94)** | 0.88 (0.80, 0.94) | 11.00 (6.32, 16.00)*† | 8.49 (5.39, 12.17)*† | **8.35 (4.85, 12.62)** | 8.94 (5.00, 13.04) | 2.52 (1.49, 3.82)*† | 1.78 (1.13, 2.67)*† | **1.73 (1.00, 2.69)** | 1.84 (1.07, 2.82) |
| Brainstem | 0.78 (0.72, 0.82)*† | 0.57 (0.48, 0.67)*† | 0.81 (0.77, 0.86) | **0.81 (0.76, 0.86)** | 2.24 (1.41, 3.00)*† | 3.16 (2.24, 5.39)*† | 2.06 (1.37, 2.79) | **2.00 (1.41, 2.83)** | 0.75 (0.57, 0.97)*† | 1.28 (0.89, 1.79)*† | **0.66 (0.48, 0.86)** | 0.70 (0.53, 0.91) |
| Cartilage(Femur) | 0.62 (0.00, 0.79)*† | 0.00 (0.00, 0.01)*† | 0.85 (0.78, 0.88) | **0.85 (0.79, 0.88)** | 28.44 (12.81, 69.89)*† | 59.91 (45.89, 72.44)*† | **8.95 (4.95, 16.07)** | 9.22 (5.10, 15.62) | 3.82 (1.82, 12.73)*† | 10.17 (8.56, 11.44)*† | **1.09 (0.86, 1.52)** | 1.10 (0.88, 1.51) |
| Cartilage(Patella) | 0.83 (0.76, 0.89)*† | 0.75 (0.64, 0.82)*† | 0.88 (0.83, 0.91) | **0.88 (0.84, 0.92)** | 4.00 (3.00, 5.10)*† | 8.94 (7.00, 10.49)*† | 3.50 (2.75, 4.95) | **3.16 (2.24, 5.00)** | 1.08 (0.71, 1.39)*† | 1.74 (1.28, 2.21)*† | 0.84 (0.65, 1.04) | **0.82 (0.61, 1.03)** |
| Cartilage(Tibia) | 0.79 (0.69, 0.84)*† | 0.59 (0.31, 0.73)*† | 0.83 (0.76, 0.88) | **0.84 (0.78, 0.88)** | 6.08 (4.12, 9.43)*† | 10.20 (8.06, 12.37)*† | **4.95 (3.50, 7.83)** | 5.00 (3.16, 7.21) | 1.08 (0.80, 1.59)*† | 2.04 (1.54, 2.96)*† | 0.86 (0.70, 1.10) | **0.84 (0.67, 1.09)** |
| Cervix Cancer | 0.81 (0.80, 0.84) | 0.88 (0.83, 0.91) | 0.82 (0.78, 0.88) | **0.86 (0.84, 0.88)** | 7.07 (3.00, 8.06) | **3.16 (3.16, 6.00)** | 6.87 (2.75, 7.58) | 5.00 (2.24, 6.40) | 1.76 (1.08, 1.81) | **0.96 (0.76, 1.24)** | 1.66 (0.99, 2.11) | 1.20 (0.84, 1.72) |
| Cochlea | 0.76 (0.70, 0.80)*† | 0.73 (0.61, 0.84)*† | **0.86 (0.80, 0.90)** | 0.86 (0.79, 0.90) | 3.61 (2.24, 5.66)*† | 5.00 (3.16, 6.40)*† | 2.75 (1.37, 3.50) | **2.24 (1.41, 4.00)** | 1.19 (0.82, 1.71)*† | 1.27 (0.85, 1.54)*† | **0.71 (0.54, 1.06)** | 0.76 (0.53, 1.12) |
| Duodenum | 0.70 (0.50, 0.82)*† | 0.62 (0.29, 0.81)*† | 0.75 (0.57, 0.89) | **0.79 (0.64, 0.90)** | 8.94 (3.04, 19.59)*† | 8.57 (4.03, 16.76)*† | 6.80 (2.75, 14.91) | **5.19 (2.38, 15.28)** | 2.09 (1.10, 4.56)*† | 2.43 (1.22, 5.14)*† | 1.77 (0.75, 4.35) | **1.49 (0.75, 3.88)** |
| Edema | 0.54 (0.35, 0.71)*† | 0.53 (0.36, 0.68)*† | 0.63 (0.43, 0.79) | **0.66 (0.51, 0.79)** | 14.87 (10.00, 19.24)*† | 13.34 (9.49, 17.26)*† | **11.73 (8.68, 15.35)** | 12.53 (9.00, 16.76) | 3.48 (2.39, 4.81)*† | 3.43 (2.63, 4.44)*† | **2.74 (1.99, 3.59)** | 2.95 (2.13, 3.88) |
| Enhancing tumour | 0.66 (0.48, 0.78)*† | 0.57 (0.40, 0.70)*† | 0.69 (0.51, 0.82) | **0.73 (0.57, 0.83)** | 10.05 (6.71, 13.67)*† | 9.90 (7.62, 12.81)*† | 9.56 (6.22, 12.43) | **9.06 (5.83, 12.08)** | 2.42 (1.66, 3.39)*† | 2.88 (2.25, 3.57)*† | 2.43 (1.57, 3.24) | **2.29 (1.39, 3.12)** |
| Esophagus | 0.84 (0.80, 0.88)*† | 0.80 (0.74, 0.86)*† | **0.93 (0.90, 0.96)** | 0.92 (0.89, 0.95) | 1.71 (1.31, 2.24)*† | 2.00 (1.41, 3.00)*† | **1.17 (0.97, 1.94)** | 1.41 (1.00, 2.00) | 0.65 (0.44, 0.99)*† | 0.71 (0.54, 0.96)*† | 0.40 (0.24, 0.57) | **0.39 (0.26, 0.54)** |
| Extra-meatal part of VS | 0.77 (0.73, 0.83)*† | 0.57 (0.47, 0.63)*† | 0.83 (0.77, 0.87) | **0.84 (0.77, 0.87)** | 2.00 (1.41, 3.00)*† | 3.61 (2.24, 5.10)*† | **1.94 (1.37, 2.17)** | 2.00 (1.41, 2.83) | 0.72 (0.57, 0.93)*† | 1.31 (0.97, 1.82)*† | **0.62 (0.43, 0.85)** | 0.65 (0.44, 0.84) |
| Gallbladder | 0.90 (0.84, 0.93)*† | 0.93 (0.89, 0.95) | 0.94 (0.91, 0.96) | **0.94 (0.91, 0.96)** | 3.08 (2.24, 3.61)*† | 2.24 (1.41, 3.12) | 2.75 (1.52, 3.50) | **2.24 (2.00, 3.12)** | 1.07 (0.82, 1.37)*† | **0.71 (0.49, 0.90)*†** | 0.87 (0.61, 1.08) | 0.81 (0.60, 1.00) |
| Heart | 0.88 (0.79, 0.94)*† | 0.82 (0.68, 0.90)*† | 0.92 (0.84, 0.96) | **0.93 (0.87, 0.96)** | 5.10 (3.16, 8.94)*† | 5.83 (2.83, 10.30)*† | 4.00 (2.17, 7.58) | **3.61 (2.24, 7.00)** | 1.34 (0.80, 1.95)*† | 1.88 (1.28, 3.06)*† | 1.00 (0.72, 1.66) | **0.97 (0.72, 1.54)** |
| Humerus cartilage | 0.62 (0.55, 0.70) | 0.64 (0.59, 0.70)*† | 0.70 (0.64, 0.75) | **0.73 (0.66, 0.79)** | 10.44 (8.00, 13.98) | 12.17 (9.22, 15.03)† | 11.40 (7.21, 18.02) | **9.85 (6.08, 15.03)** | 3.17 (2.49, 3.73) | 3.05 (2.43, 3.57)*† | 2.64 (2.03, 3.69) | **2.15 (1.55, 3.37)** |
| Inferior Vena Cava | 0.90 (0.87, 0.93)*† | 0.95 (0.92, 0.96) | **0.95 (0.92, 0.96)** | 0.94 (0.91, 0.96) | 2.24 (1.41, 3.00)*† | **1.41 (1.00, 2.00)*†** | 1.94 (1.37, 2.17) | 2.00 (1.41, 2.24) | 0.76 (0.56, 0.98)*† | **0.44 (0.32, 0.58)*†** | 0.54 (0.39, 0.80) | 0.58 (0.40, 0.83) |
| Ischemic Stroke | 0.79 (0.69, 0.86)*† | **0.86 (0.76, 0.93)*†** | 0.83 (0.74, 0.90) | 0.84 (0.75, 0.90) | 3.16 (2.00, 6.00)*† | **2.83 (2.00, 4.03)*†** | 3.07 (1.94, 4.95) | 3.16 (2.00, 5.66) | 0.94 (0.64, 1.43)*† | **0.63 (0.44, 0.88)*†** | 0.85 (0.53, 1.29) | 0.85 (0.56, 1.35) |
| Kidney | 0.95 (0.93, 0.96) | 0.95 (0.94, 0.96) | 0.98 (0.98, 0.99) | **0.98 (0.97, 1.00)** | 19.53 (18.35, 23.44) | 9.18 (4.68, 13.10) | 5.43 (4.34, 8.80) | **4.47 (4.21, 8.23)** | 1.94 (1.65, 2.61) | 1.82 (1.35, 2.05) | 1.37 (1.24, 1.39) | **1.25 (0.99, 1.37)** |
| Left Adrenal Gland | 0.77 (0.72, 0.82)*† | 0.63 (0.56, 0.69)*† | 0.84 (0.77, 0.88) | **0.85 (0.77, 0.88)** | 2.00 (1.41, 2.83)*† | 3.16 (2.24, 4.74)*† | **1.94 (1.37, 2.17)** | 2.00 (1.41, 2.24) | 0.75 (0.60, 0.98)*† | 1.00 (0.78, 1.27)*† | **0.54 (0.41, 0.77)** | 0.58 (0.42, 0.81) |
| Left Kidney | 0.94 (0.89, 0.95)*† | 0.96 (0.93, 0.97) | **0.96 (0.94, 0.98)** | 0.95 (0.93, 0.97) | 4.00 (2.24, 12.17)*† | **2.83 (1.41, 4.47)*†** | 3.88 (1.94, 7.83) | 4.00 (2.00, 9.14) | 1.20 (0.72, 1.80)*† | **0.60 (0.39, 1.04)*†** | 0.89 (0.63, 1.45) | 1.02 (0.61, 1.61) |
| Left Ventricle | **0.55 (0.40, 0.71)*†** | 0.05 (0.02, 0.26)*† | 0.27 (0.03, 0.68) | 0.49 (0.27, 0.71) | **6.40 (5.00, 8.74)*†** | 7.00 (5.83, 8.06)*† | 7.39 (5.66, 9.16) | 7.21 (5.39, 10.30) | 1.61 (1.30, 1.94)*† | 2.57 (2.13, 2.88)*† | 2.29 (1.80, 2.79) | **1.98 (1.55, 2.46)** |
| Liver | 0.93 (0.89, 0.95)*† | **0.97 (0.93, 0.98)*†** | 0.95 (0.91, 0.97) | 0.95 (0.91, 0.97) | 14.32 (4.47, 27.46)*† | **5.83 (3.00, 13.00)*†** | 12.62 (4.34, 23.16) | 12.81 (4.47, 22.56) | 2.18 (1.15, 3.84)*† | **0.94 (0.52, 2.02)*†** | 1.89 (1.21, 3.55) | 1.99 (1.13, 3.34) |
| Myocardium | **0.95 (0.90, 0.97)*†** | 0.94 (0.90, 0.96)*† | 0.91 (0.84, 0.95) | 0.93 (0.89, 0.95) | 3.00 (2.00, 4.47)*† | **2.83 (2.00, 4.12)*†** | 4.85 (3.07, 7.92) | 4.12 (2.83, 6.40) | **0.66 (0.45, 1.00)*†** | 0.72 (0.56, 0.95)*† | 1.27 (0.83, 1.97) | 1.02 (0.73, 1.39) |
| Non-enhancing tumor | 0.48 (0.28, 0.68)*† | 0.46 (0.24, 0.65)*† | 0.50 (0.30, 0.70) | **0.54 (0.33, 0.74)** | 11.05 (7.28, 15.65)*† | 10.63 (8.06, 14.00)*† | 10.32 (7.39, 13.54) | **10.20 (7.21, 13.45)** | **3.10 (2.10, 4.26)*†** | 3.14 (2.44, 4.00)*† | 2.98 (2.08, 3.92) | 2.89 (2.02, 3.79) |
| Pancreas | 0.83 (0.72, 0.89)*† | **0.91 (0.83, 0.93)*†** | 0.88 (0.78, 0.92) | 0.88 (0.79, 0.93) | 7.62 (3.16, 13.82)*† | **4.24 (2.24, 7.45)*†** | 5.91 (2.75, 10.39) | 5.66 (2.83, 11.40) | 1.52 (0.97, 2.84)*† | **0.96 (0.65, 1.49)*†** | 1.32 (0.82, 2.42) | 1.27 (0.78, 2.59) |
| Patella | 0.88 (0.83, 0.91)*† | 0.85 (0.80, 0.89)*† | 0.94 (0.85, 0.96) | **0.95 (0.88, 0.96)** | 6.71 (5.00, 9.43)*† | 8.49 (6.40, 10.44)*† | **4.85 (3.88, 7.77)** | 5.00 (4.00, 7.00) | 1.93 (1.43, 2.55)*† | 2.43 (1.90, 3.04)*† | 1.36 (1.06, 1.82) | **1.24 (0.99, 1.63)** |
| Posterior Hippocampus | 0.83 (0.77, 0.86)*† | 0.73 (0.65, 0.79)*† | **0.90 (0.85, 0.92)** | 0.89 (0.86, 0.92) | 2.24 (2.00, 3.16)*† | 3.00 (2.83, 3.61)*† | **1.94 (1.37, 2.17)** | 2.00 (1.41, 2.24) | 0.74 (0.60, 0.92)*† | 1.05 (0.89, 1.26)*† | **0.53 (0.40, 0.66)** | 0.55 (0.41, 0.68) |
| Prostate | 0.91 (0.87, 0.94)*† | 0.93 (0.87, 0.96)*† | 0.95 (0.91, 0.97) | **0.95 (0.92, 0.97)** | 9.85 (5.10, 15.52)*† | 6.71 (3.00, 12.21)*† | 6.80 (4.00, 9.90) | **6.40 (4.12, 9.85)** | 2.47 (1.45, 3.89)*† | 2.19 (1.01, 4.50)*† | 2.07 (1.31, 3.23) | **2.03 (1.30, 3.04)** |
| Prostate Central Zone | 0.89 (0.85, 0.92)* | 0.78 (0.63, 0.87)*† | 0.88 (0.76, 0.93) | **0.91 (0.84, 0.93)** | 8.00 (5.66, 10.77)† | 9.85 (7.62, 12.00)*† | 8.12 (5.83, 11.07) | **6.32 (5.10, 9.01)** | **2.27 (1.67, 2.95)*†** | 4.01 (2.80, 5.41)*† | 3.15 (1.82, 4.92) | 2.45 (1.89, 3.60) |
| Prostate Peripheral Zone | 0.61 (0.51, 0.72)† | 0.44 (0.30, 0.54)*† | 0.63 (0.48, 0.78) | **0.73 (0.61, 0.82)** | 13.42 (6.40, 21.95)*† | 16.55 (8.94, 25.32)*† | 9.90 (5.23, 20.39) | **9.00 (5.00, 16.00)** | 3.48 (1.74, 5.73)† | 4.44 (2.65, 6.50)*† | 2.88 (1.50, 5.43) | **2.36 (1.47, 3.93)** |
| Rectum | 0.91 (0.87, 0.94)*† | 0.78 (0.64, 0.85)*† | **0.93 (0.89, 0.96)** | 0.93 (0.89, 0.95) | 2.83 (2.24, 4.12)*† | 5.83 (4.24, 7.36)*† | **2.75 (1.94, 4.00)** | 3.00 (2.00, 4.00) | **0.86 (0.70, 1.22)** | 2.24 (1.58, 3.04)*† | 1.01 (0.60, 1.48) | 1.02 (0.71, 1.39) |
| Right Adrenal Gland | 0.69 (0.60, 0.78)*† | 0.65 (0.55, 0.72)*† | 0.76 (0.73, 0.81) | **0.78 (0.74, 0.83)** | 2.24 (1.41, 3.00)† | 3.16 (2.24, 4.24)*† | **1.94 (1.94, 2.75)** | 2.00 (1.41, 2.24) | 0.78 (0.56, 1.10)*† | 0.85 (0.69, 1.09)*† | **0.70 (0.56, 0.78)** | 0.71 (0.53, 0.84) |
| Right Kidney | 0.93 (0.90, 0.96)*† | **0.96 (0.92, 0.97)*†** | 0.95 (0.93, 0.96) | 0.94 (0.92, 0.96) | 4.00 (2.24, 11.66)*† | **3.00 (1.41, 5.00)*†** | 3.50 (2.17, 7.58) | 3.61 (2.24, 10.00) | 1.09 (0.72, 1.67) | **0.65 (0.40, 1.12)*†** | 0.98 (0.78, 1.40) | 0.99 (0.74, 1.53) |
| Right Ventricle | 0.94 (0.90, 0.97)* | **0.95 (0.92, 0.97)*†** | 0.93 (0.88, 0.96) | 0.94 (0.91, 0.96) | 2.83 (2.00, 4.24)*† | **2.00 (1.41, 2.24)*†** | 3.50 (1.94, 5.91) | 3.00 (2.00, 5.00) | 0.71 (0.49, 1.18)*† | **0.66 (0.53, 0.85)*†** | 1.09 (0.66, 1.93) | 0.96 (0.59, 1.50) |
| Scapula cartilage | 0.01 (0.00, 0.14) | 0.00 (0.00, 0.01)*† | 0.48 (0.08, 0.61) | **0.66 (0.60, 0.74)** | 92.00 (56.58, 116.09) | 84.81 (63.16, 100.12)*† | 41.59 (21.14, 96.84) | **22.59 (13.01, 38.08)** | 20.18 (11.35, 24.91) | 18.02 (14.13, 21.16)*† | 6.75 (4.41, 19.45) | **3.42 (2.50, 4.86)** |
| Spleen | 0.94 (0.90, 0.96)*† | 0.96 (0.92, 0.98)† | 0.95 (0.92, 0.98) | **0.96 (0.93, 0.98)** | 3.00 (2.24, 5.07)† | **2.83 (1.41, 4.47)*†** | 2.91 (1.94, 5.23) | 3.00 (2.00, 5.00) | 0.96 (0.63, 1.27)* | **0.67 (0.42, 1.13)*†** | 0.90 (0.66, 1.41) | 0.86 (0.69, 1.28) |
| Stomach | 0.83 (0.72, 0.89)*† | **0.91 (0.83, 0.95)*†** | 0.87 (0.79, 0.92) | 0.88 (0.83, 0.92) | 10.79 (7.02, 16.70)*† | **6.00 (4.00, 9.85)*†** | 9.56 (5.66, 14.43) | 8.97 (6.00, 14.50) | 2.69 (1.83, 3.79)*† | **1.24 (0.89, 2.14)*†** | 2.24 (1.64, 3.95) | 2.35 (1.55, 3.69) |
| Total body MRI | 0.86 (0.78, 0.91)*† | 0.78 (0.62, 0.88)*† | 0.91 (0.82, 0.95) | **0.91 (0.84, 0.95)** | 7.07 (3.16, 15.62)*† | 9.22 (5.00, 17.49)*† | 6.14 (2.75, 14.56) | **5.83 (2.83, 13.60)** | 1.84 (1.05, 3.42)*† | 2.80 (1.55, 4.92)*† | **1.56 (0.79, 3.44)** | 1.57 (0.82, 3.19) |
| Vertebrae | 0.87 (0.76, 0.91)*† | 0.92 (0.88, 0.94)*† | **0.92 (0.89, 0.94)** | 0.92 (0.88, 0.94) | 12.65 (6.71, 47.75)*† | **7.28 (4.47, 16.28)*†** | 8.68 (4.85, 18.68) | 9.43 (5.00, 19.42) | 2.75 (1.73, 9.04)*† | **1.88 (1.33, 2.80)*†** | 2.09 (1.48, 3.34) | 2.27 (1.58, 3.53) |
| Vestibular Schwannoma | 0.90 (0.85, 0.95)*† | 0.93 (0.87, 0.97)*† | **0.95 (0.92, 0.97)** | 0.94 (0.92, 0.97) | 2.83 (2.00, 4.74)*† | 2.24 (1.41, 5.00)*† | **2.17 (1.37, 2.91)** | 2.24 (1.41, 3.00) | 0.82 (0.57, 1.23)*† | **0.60 (0.36, 1.00)** | 0.64 (0.46, 0.89) | 0.66 (0.47, 0.94) |
| White Matter Hyperintensity | 0.74 (0.56, 0.84)*† | 0.66 (0.48, 0.78)*† | 0.86 (0.74, 0.92) | **0.86 (0.76, 0.92)** | 2.83 (1.41, 8.00)*† | 5.00 (2.83, 8.25)*† | 2.17 (0.97, 6.22) | **2.00 (1.00, 6.00)** | 0.88 (0.55, 1.84)*† | 1.18 (0.78, 2.02)*† | **0.51 (0.31, 1.07)** | 0.53 (0.30, 1.15) |
| AVG | 0.80 ± 0.17 | 0.74 ± 0.24 | 0.85 ± 0.15 | **0.87 ± 0.11** | 8.61 ± 13.56 | 8.89 ± 14.23 | 5.95 ± 6.21 | **5.36 ± 4.07** | 1.99 ± 2.85 | 2.20 ± 2.84 | 1.42 ± 1.10 | **1.30 ± 0.76** |

Scores are reported as median (Q1, Q3). “Average” denotes mean ± SD of the medians across tasks. * indicates **SAMRI$_{box}$** is significantly better than both other models ($p < 0.05$); † indicates **SAMRI$_{bp}$** is significantly better than both other models ($p < 0.05$). **SAMRI$_{box}$** = SAMRI with box-only prompt; **SAMRI$_{bp}$** = SAMRI with box+point prompt.

**Table S4: Quantitative performance of SAMRI compared with SAM vit-b and MedSAM across 36 Datasets (DSC, HD, and MSD)**

| Dataset Name | Dice↑ | | | | HD↓ | | | | MSD↓ | | | |
|---|---|---|---|---|---|---|---|---|---|---|---|---|
| | SAM- ViTB | MedSAM | SAMRI$_{box}$ | SAMRI$_{bp}$ | SAM- ViTB | MedSAM | SAMRI$_{box}$ | SAMRI$_{bp}$ | SAM- ViTB | MedSAM | SAMRI$_{box}$ | SAMRI$_{bp}$ |
| ACDC# | 0.90 (0.66, 0.96)*† | 0.90 (0.22, 0.95)*† | 0.87 (0.55, 0.94) | 0.90 (0.66, 0.95) | 4.00 (2.24, 6.40)*† | 3.16 (2.00, 6.32)*† | 5.66 (3.07, 7.83) | 5.00 (2.83, 7.62) | 1.01 (0.58, 1.58)*† | 0.92 (0.62, 2.25)* | 1.62 (0.93, 2.43) | 1.29 (0.81, 1.95) |
| AMOSMR | 0.90 (0.83, 0.94)*† | 0.95 (0.88, 0.97)*† | 0.94 (0.89, 0.97) | 0.94 (0.89, 0.97) | 3.16 (2.00, 10.00)*† | 2.24 (1.00, 5.07)*† | 2.75 (1.37, 7.77) | 2.83 (1.41, 8.00) | 0.99 (0.64, 1.95)*† | 0.59 (0.36, 1.05)*† | 0.81 (0.44, 1.63) | 0.82 (0.46, 1.66) |
| Brain_Tumor_Dataset_Figshare | 0.92 (0.86, 0.95)*† | 0.80 (0.71, 0.87)*† | 0.94 (0.91, 0.97) | 0.94 (0.91, 0.96) | 7.62 (5.00, 14.71)*† | 14.76 (11.18, 19.08)*† | 6.22 (4.12, 11.57) | 6.71 (4.47, 11.54) | 2.09 (1.51, 3.53)*† | 5.44 (4.01, 7.25)*† | 1.93 (1.36, 3.08) | 2.03 (1.46, 3.19) |
| BraTS_FLAIR | 0.85 (0.75, 0.90)*† | 0.91 (0.83, 0.95)*† | 0.92 (0.84, 0.96) | 0.91 (0.84, 0.95) | 11.31 (7.07, 16.12)*† | 7.62 (4.47, 11.66)*† | 8.68 (4.95, 12.62) | 9.06 (5.39, 13.04) | 2.34 (1.53, 3.40)*† | 1.30 (0.86, 1.97)*† | 1.47 (1.00, 2.20) | 1.56 (1.05, 2.31) |
| BraTS_T1 | 0.77 (0.64, 0.84)*† | 0.84 (0.73, 0.90)*† | 0.85 (0.75, 0.90) | 0.85 (0.76, 0.90) | 13.00 (8.25, 17.69)*† | 9.22 (6.40, 13.00)*† | 10.32 (6.87, 13.80) | 10.77 (7.07, 14.32) | 3.27 (2.22, 4.52)*† | 2.14 (1.58, 2.90)*† | 2.47 (1.78, 3.26) | 2.57 (1.80, 3.44) |
| BraTS_T1CE | 0.82 (0.63, 0.93)*† | 0.72 (0.52, 0.86)*† | 0.89 (0.78, 0.94) | 0.88 (0.77, 0.94) | 5.83 (3.00, 10.00)*† | 7.07 (4.12, 10.00)*† | 4.34 (2.17, 7.58) | 4.47 (2.24, 8.06) | 1.18 (0.58, 2.45)*† | 1.71 (1.05, 2.69)*† | 0.77 (0.54, 1.20) | 0.87 (0.57, 1.50) |
| CervicalCancer# | 0.81 (0.80, 0.84) | 0.88 (0.83, 0.91) | 0.82 (0.78, 0.88) | 0.86 (0.84, 0.88) | 7.07 (3.00, 8.06) | 3.16 (3.16, 6.00) | 6.87 (2.75, 7.58) | 5.00 (2.24, 6.40) | 1.76 (1.08, 1.81) | 0.96 (0.76, 1.24) | 1.66 (0.99, 2.11) | 1.20 (0.84, 1.72) |
| CHAOS_T1# | 0.92 (0.89, 0.95)† | 0.92 (0.88, 0.95)† | 0.93 (0.89, 0.95) | 0.94 (0.91, 0.96) | 3.61 (2.83, 7.07)* | 4.00 (3.00, 5.83)*† | 4.00 (2.87, 8.46) | 4.06 (2.24, 7.21) | 1.24 (0.93, 1.76)* | 1.39 (0.89, 1.90)* | 1.29 (0.97, 2.04) | 1.18 (0.84, 1.73) |
| CHAOS_T2# | 0.94 (0.89, 0.96)* | 0.88 (0.80, 0.92)† | 0.93 (0.85, 0.96) | 0.94 (0.90, 0.96) | 4.36 (2.24, 12.29)* | 6.56 (4.24, 10.81)* | 5.36 (2.31, 15.31) | 5.00 (2.24, 11.03) | 1.01 (0.65, 1.95)*† | 2.14 (1.36, 2.99)† | 1.35 (0.79, 3.55) | 1.09 (0.79, 2.22) |
| crossmoda | 0.82 (0.75, 0.90)*† | 0.73 (0.58, 0.92)*† | 0.89 (0.81, 0.94) | 0.88 (0.81, 0.94) | 2.83 (2.00, 3.61)*† | 3.16 (2.00, 5.39)*† | 2.17 (1.37, 2.91) | 2.24 (1.41, 3.00) | 0.82 (0.60, 1.19)*† | 0.99 (0.62, 1.51)*† | 0.65 (0.47, 0.90) | 0.69 (0.48, 0.95) |
| Heart | 0.89 (0.79, 0.95)† | 0.90 (0.84, 0.92)† | 0.90 (0.81, 0.95) | 0.92 (0.86, 0.96) | 5.00 (3.16, 8.06)† | 2.83 (2.24, 4.30)* | 5.49 (2.91, 9.39) | 4.24 (2.83, 7.14) | 1.14 (0.75, 1.86)† | 1.29 (1.16, 1.47)† | 1.21 (0.84, 2.24) | 1.05 (0.78, 1.66) |
| HipMRI | 0.86 (0.81, 0.90)*† | 0.73 (0.58, 0.82)*† | 0.95 (0.92, 0.97) | 0.94 (0.91, 0.96) | 3.61 (2.24, 5.00)*† | 6.08 (4.24, 8.00)*† | 2.17 (1.37, 3.88) | 2.24 (1.41, 4.00) | 1.41 (0.94, 1.85)*† | 2.41 (1.74, 3.33)*† | 0.74 (0.51, 1.04) | 0.77 (0.53, 1.09) |
| ISLES2022_ADC | 0.76 (0.67, 0.84)*† | 0.83 (0.72, 0.92)*† | 0.80 (0.70, 0.87) | 0.81 (0.72, 0.87) | 3.61 (2.24, 6.63)*† | 3.00 (2.00, 4.24)*† | 3.88 (2.17, 5.49) | 3.61 (2.00, 5.96) | 1.03 (0.77, 1.57)*† | 0.73 (0.54, 0.99)*† | 0.99 (0.67, 1.41) | 1.01 (0.68, 1.54) |
| ISLES2022_DWI | 0.84 (0.74, 0.89)*† | 0.88 (0.82, 0.94) | 0.87 (0.82, 0.92) | 0.87 (0.81, 0.92) | 3.00 (1.41, 5.00)*† | 2.00 (1.41, 3.61)*† | 2.17 (1.37, 4.73) | 2.83 (1.41, 4.47) | 0.74 (0.45, 1.23)*† | 0.48 (0.35, 0.68)*† | 0.60 (0.40, 1.00) | 0.63 (0.42, 1.07) |
| MSD_BrainTumour | 0.56 (0.36, 0.73)*† | 0.52 (0.34, 0.68)*† | 0.62 (0.41, 0.78) | 0.66 (0.47, 0.79) | 12.08 (8.06, 17.03)*† | 11.40 (8.49, 15.35)*† | 10.68 (7.58, 14.14) | 11.00 (7.62, 14.76) | 3.06 (2.06, 4.32)*† | 3.18 (2.46, 4.08)*† | 2.71 (1.91, 3.60) | 2.76 (1.89, 3.68) |
| MSD_Heart | 0.87 (0.79, 0.93)*† | 0.71 (0.62, 0.79)*† | 0.93 (0.87, 0.97) | 0.93 (0.88, 0.97) | 5.39 (3.61, 9.16)*† | 9.00 (6.32, 10.82)*† | 2.99 (1.94, 5.83) | 3.08 (2.24, 6.00) | 1.48 (0.94, 2.02)*† | 3.05 (2.30, 3.94)*† | 0.89 (0.66, 1.27) | 0.92 (0.69, 1.29) |
| MSD_Hippocampus | 0.83 (0.77, 0.86)*† | 0.75 (0.67, 0.80)*† | 0.90 (0.86, 0.93) | 0.90 (0.86, 0.93) | 2.83 (2.00, 3.61)*† | 3.16 (2.83, 4.00)*† | 1.94 (1.37, 2.17) | 2.00 (1.41, 2.24) | 0.81 (0.64, 1.01)*† | 1.08 (0.91, 1.32)*† | 0.54 (0.42, 0.69) | 0.58 (0.44, 0.72) |
| MSD_Prostate | 0.85 (0.63, 0.90)*† | 0.69 (0.46, 0.84)*† | 0.81 (0.65, 0.90) | 0.86 (0.73, 0.92) | 10.00 (7.00, 14.12)† | 11.36 (8.60, 15.47)*† | 9.36 (6.58, 13.90) | 7.71 (5.83, 11.40) | 2.75 (2.01, 3.64)* | 4.47 (3.16, 5.81)*† | 3.32 (2.02, 5.38) | 2.59 (1.95, 3.92) |
| MSK_FLASH | 0.87 (0.75, 0.93)*† | 0.82 (0.47, 0.91)*† | 0.91 (0.82, 0.96) | 0.91 (0.83, 0.96) | 9.49 (5.66, 16.51)*† | 12.04 (8.60, 20.00)*† | 7.77 (4.85, 13.03) | 7.81 (5.00, 12.65) | 2.15 (1.47, 3.38)*† | 3.13 (2.23, 4.89)*† | 1.57 (1.07, 2.48) | 1.52 (1.04, 2.39) |
| MSK_PD | 0.86 (0.72, 0.93)*† | 0.82 (0.45, 0.91)*† | 0.88 (0.78, 0.95) | 0.89 (0.79, 0.95) | 9.43 (5.00, 16.12)*† | 12.21 (9.00, 21.01)*† | 8.01 (4.85, 14.96) | 8.06 (4.47, 14.21) | 2.08 (1.27, 3.03)*† | 3.25 (2.14, 5.28)*† | 1.66 (0.98, 3.20) | 1.59 (0.96, 2.82) |
| MSK_shoulder# | 0.27 (0.00, 0.62) | 0.02 (0.00, 0.62)*† | 0.59 (0.30, 0.70) | 0.69 (0.61, 0.75) | 47.68 (11.66, 103.17) | 58.00 (13.60, 90.14)*† | 22.63 (12.37, 53.67) | 16.22 (9.22, 28.30) | 10.08 (3.20, 22.11) | 13.46 (3.28, 19.50)*† | 4.70 (2.84, 11.04) | 3.03 (2.07, 4.35) |
| MSK_T2 | 0.90 (0.82, 0.94)*† | 0.86 (0.59, 0.93)*† | 0.92 (0.86, 0.97) | 0.93 (0.87, 0.97) | 8.60 (5.00, 16.00)*† | 10.63 (7.62, 18.11)*† | 5.91 (3.88, 9.71) | 6.08 (4.00, 10.20) | 1.70 (1.03, 2.86)*† | 2.82 (1.76, 4.58)*† | 1.16 (0.76, 1.90) | 1.13 (0.74, 1.89) |
| OAIAKOA | 0.88 (0.71, 0.96)*† | 0.82 (0.34, 0.95)*† | 0.91 (0.84, 0.99) | 0.92 (0.84, 0.99) | 8.00 (4.47, 18.30)*† | 12.04 (8.00, 24.76)*† | 4.95 (3.50, 7.83) | 5.10 (3.61, 8.06) | 1.48 (1.02, 2.81)*† | 2.36 (1.64, 5.32)*† | 0.96 (0.78, 1.24) | 1.00 (0.80, 1.28) |
| Picai | 0.91 (0.86, 0.95)*† | 0.89 (0.82, 0.93)*† | 0.96 (0.93, 0.98) | 0.96 (0.93, 0.98) | 13.00 (8.94, 18.68)*† | 11.68 (8.06, 16.97)*† | 6.87 (4.85, 9.76) | 7.21 (5.00, 10.00) | 3.16 (2.19, 4.69)*† | 4.24 (2.84, 6.02)*† | 2.18 (1.59, 3.09) | 2.19 (1.60, 3.16) |
| PROMISE | 0.94 (0.90, 0.96)† | 0.86 (0.79, 0.91)*† | 0.94 (0.88, 0.97) | 0.94 (0.90, 0.96) | 8.06 (5.00, 12.00)† | 11.25 (8.06, 13.68)*† | 8.24 (5.66, 10.69) | 8.00 (5.00, 11.01) | 2.10 (1.34, 3.09)*† | 4.13 (2.79, 5.18)*† | 2.65 (1.77, 4.00) | 2.65 (1.68, 3.63) |
| ProstateADC | 0.90 (0.87, 0.92)*† | 0.95 (0.92, 0.96)* | 0.94 (0.90, 0.96) | 0.94 (0.92, 0.96) | 3.16 (2.83, 4.24)*† | 2.00 (1.00, 2.83)*† | 2.17 (1.94, 3.50) | 2.24 (2.00, 3.00) | 1.05 (0.82, 1.33)*† | 0.61 (0.44, 0.89)*† | 0.78 (0.59, 1.16) | 0.75 (0.58, 1.02) |
| ProstateT2 | 0.93 (0.89, 0.95)*† | 0.96 (0.95, 0.97)*† | 0.94 (0.90, 0.97) | 0.95 (0.91, 0.97) | 9.43 (6.40, 14.34)*† | 4.00 (3.00, 5.00)*† | 7.39 (4.95, 10.76) | 7.00 (5.00, 9.85) | 2.39 (1.66, 3.35)† | 1.16 (0.94, 1.54)*† | 2.35 (1.60, 3.34) | 2.14 (1.63, 2.99) |
| QIN-PROSTATE-Lesion | 0.76 (0.69, 0.79)*† | 0.40 (0.15, 0.56)*† | 0.82 (0.72, 0.86) | 0.85 (0.77, 0.89) | 2.00 (1.41, 3.16)*† | 4.47 (3.61, 6.40)*† | 1.94 (1.37, 2.17) | 2.00 (1.41, 2.24) | 0.69 (0.55, 1.11)*† | 1.58 (1.35, 2.61)*† | 0.63 (0.48, 0.76) | 0.59 (0.37, 0.77) |
| QIN-PROSTATE-Prostate | 0.89 (0.80, 0.92)*† | 0.96 (0.95, 0.97)*† | 0.92 (0.87, 0.96) | 0.94 (0.90, 0.96) | 11.00 (7.00, 18.00)*† | 4.24 (2.83, 7.62)*† | 8.68 (6.51, 11.69) | 8.06 (5.00, 10.63) | 3.59 (2.16, 5.57)*† | 1.17 (0.79, 2.27)*† | 2.91 (2.05, 3.81) | 2.37 (1.72, 3.68) |
| QUBIQ_kidney | 0.95 (0.93, 0.96) | 0.95 (0.94, 0.96) | 0.98 (0.98, 0.99) | 0.98 (0.97, 1.00) | 19.53 (18.35, 23.44) | 9.18 (4.68, 13.10) | 5.43 (4.34, 8.80) | 4.47 (4.21, 8.23) | 1.94 (1.65, 2.61) | 1.82 (1.35, 2.05) | 1.37 (1.24, 1.39) | 1.25 (0.99, 1.37) |
| QUBIQ_prostate# | 0.93 (0.90, 0.96)* | 0.89 (0.87, 0.92)† | 0.91 (0.87, 0.96) | 0.94 (0.91, 0.96) | 27.27 (14.76, 41.03) | 32.41 (18.90, 44.61) | 23.58 (20.30, 38.23) | 22.84 (16.41, 33.11) | 5.53 (3.93, 8.74)* | 8.48 (7.19, 14.17)† | 7.63 (6.45, 15.62) | 7.28 (4.60, 11.13) |
| SpineMR | 0.87 (0.76, 0.91)*† | 0.92 (0.88, 0.94)*† | 0.92 (0.89, 0.94) | 0.92 (0.88, 0.94) | 12.65 (6.71, 47.75)*† | 7.28 (4.47, 16.28)*† | 8.68 (4.85, 18.68) | 9.43 (5.00, 19.42) | 2.75 (1.73, 9.04)*† | 1.88 (1.33, 2.80)*† | 2.09 (1.48, 3.34) | 2.27 (1.58, 3.53) |
| totalseg_mr | 0.86 (0.78, 0.91)*† | 0.78 (0.62, 0.88)*† | 0.91 (0.82, 0.95) | 0.91 (0.84, 0.95) | 7.07 (3.16, 15.62)*† | 9.22 (5.00, 17.49)*† | 6.14 (2.75, 14.56) | 5.83 (2.83, 13.60) | 1.84 (1.05, 3.42)*† | 2.80 (1.55, 4.92)*† | 1.56 (0.79, 3.44) | 1.57 (0.82, 3.19) |
| WMH_FLAIR | 0.79 (0.64, 0.87)*† | 0.69 (0.49, 0.82)*† | 0.90 (0.83, 0.94) | 0.89 (0.82, 0.94) | 2.24 (1.41, 8.37)*† | 5.00 (2.24, 9.03)*† | 1.37 (0.97, 5.91) | 1.41 (1.00, 5.52) | 0.76 (0.47, 1.55)*† | 1.08 (0.68, 2.09)*† | 0.40 (0.24, 0.78) | 0.41 (0.25, 0.77) |
| WMH_T1 | 0.69 (0.48, 0.79)*† | 0.61 (0.48, 0.73)*† | 0.79 (0.67, 0.87) | 0.81 (0.69, 0.89) | 3.00 (2.00, 7.28)*† | 5.00 (3.00, 7.62)*† | 2.75 (1.37, 6.80) | 2.24 (1.41, 6.08) | 1.01 (0.63, 2.00)*† | 1.28 (0.89, 1.97)*† | 0.72 (0.41, 1.33) | 0.71 (0.41, 1.40) |
| ZIB_OAI | 0.91 (0.74, 0.97)*† | 0.89 (0.04, 0.96)*† | 0.95 (0.85, 1.00) | 0.95 (0.85, 0.99) | 9.22 (5.00, 23.00)*† | 12.00 (7.62, 31.98)*† | 6.14 (4.00, 9.90) | 6.40 (4.12, 10.44) | 1.59 (1.00, 3.33)*† | 2.56 (1.67, 6.61)*† | 1.06 (0.83, 1.41) | 1.10 (0.86, 1.48) |
| AVG | 0.84 ± 0.13 | 0.80 ± 0.18 | 0.89 ± 0.08 | 0.90 ± 0.07 | 8.86 ± 8.45 | 9.23 ± 10.07 | 6.49 ± 4.85 | 6.17 ± 4.27 | 2.06 ± 1.71 | 2.56 ± 2.47 | 1.71 ± 1.37 | 1.59 ± 1.22 |

Datasets marked with '#' are zero-shot. "Average" denotes mean ± SD of the medians across tasks. * indicates **SAMRI$_{box}$** is significantly better than both other models ($p < 0.05$); † indicates **SAMRI$_{bp}$** is significantly better than both other models ($p < 0.05$). **SAMRI$_{box}$** = SAMRI with box-only prompt; **SAMRI$_{bp}$** = SAMRI with box+point prompt.

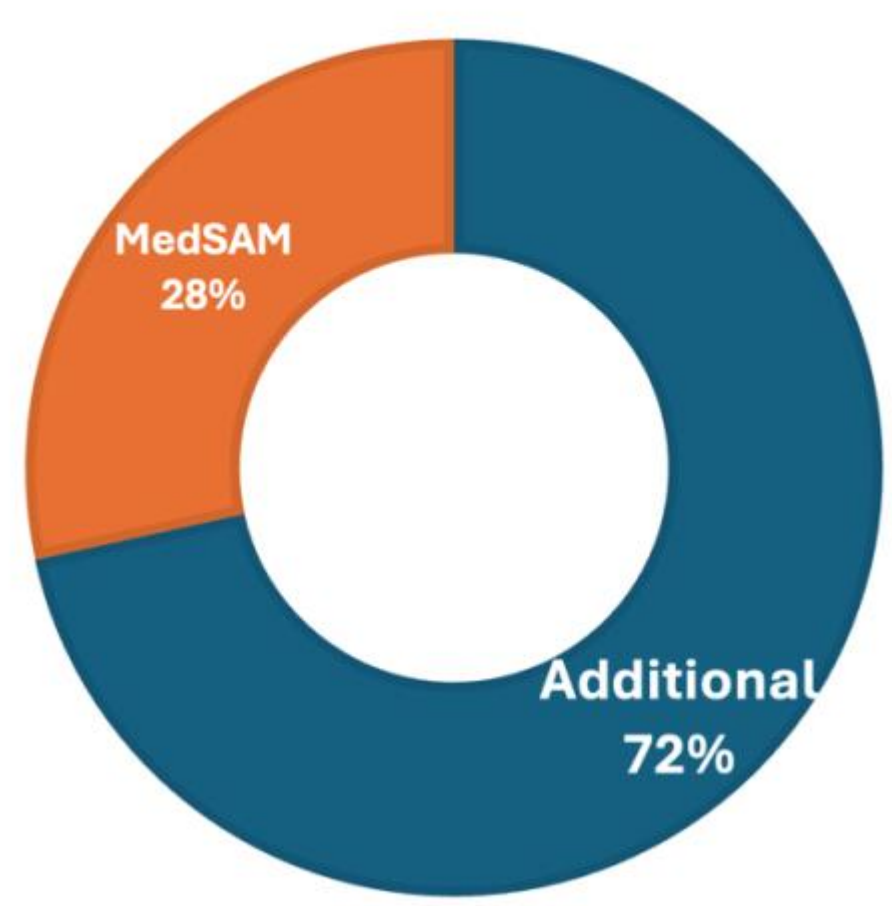

**Fig. S1** *Dataset expansion beyond MedSAM.* We inherited **21** MRI datasets from MedSAM and added **15** new datasets covering additional tasks and sequences, **tripling** the MRI sample count relative to MedSAM. Consequently, in the final training corpus, **~72%** of images come from the newly added datasets, while **~28%** originate from MedSAM's collection. This expansion broadens anatomical and sequence diversity, supporting stronger zero-shot generalization.

**Table S5. Computational comparison between SAM components**

| Components | Parameters | Inference time |
|---|---|---|
| Image Encoder | 89.67M | 0.15s (GPU) |
| Mask Decoder | 4M | 0.05s (CPU) |

**Table S6 Experimental results on a Mac mini (Apple M4 Pro; 24 GB unified memory).**

| Project | Image →mask (precomputing) | Image →mask (Inference) | Train from embedding (Embedding once) | Train from image (Embedding every epoch) |
|---|---|---|---|---|
| Time | 50.43 seconds | 55.09 seconds | 4.42 seconds/epoch | 63.23 second/epoch |
| Memory usage | 4.40 GB | 4.46 GB | 2.28 GB | 6.73 GB |

Note: The results are based on 100 2D NIfTI-format MR image-mask pairs.

**Table S7 Training resources and total GPU hours across models.**

| Model | Training Resource* | Total GPU hours[n] |
|---|---|---|
| SAM ViT-B | ~5 days @256 A100 | ~10,000 |
| MedSAM | ~21 days @20 A100 | ~10,000 |
| **SAMRI$_{hox}$ (box)** | **~3 days @8 MI300X** | **608** |
| **SAMRI$_{ho}$ (box+point)** | **~3 days @8 MI300X** | **608** |

* Based on the original training dataset. [n] Standardised by 1× GPU / 1M training samples / 100 epochs.
SAMRI$_{hox}$ = box-only prompts; SAMRI$_{ho}$ = box+point prompts.

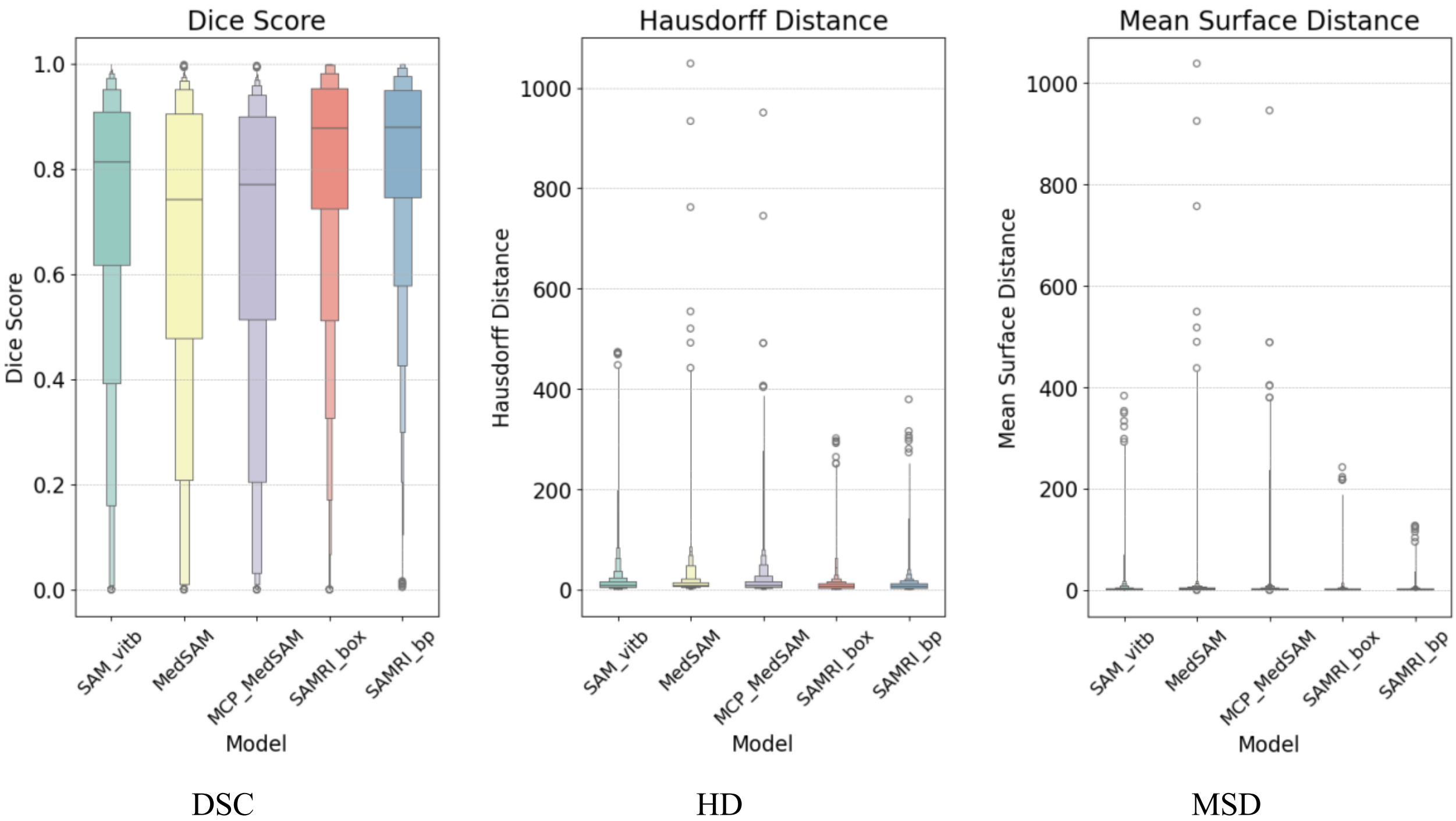

**Fig. S2**: *The overall segmentation performance for five models*. $\mathbf{SAMRI_{box}}$ = SAMRI with box-only prompt; $\mathbf{SAMRI_{bp}}$ = SAMRI with box+point prompt.

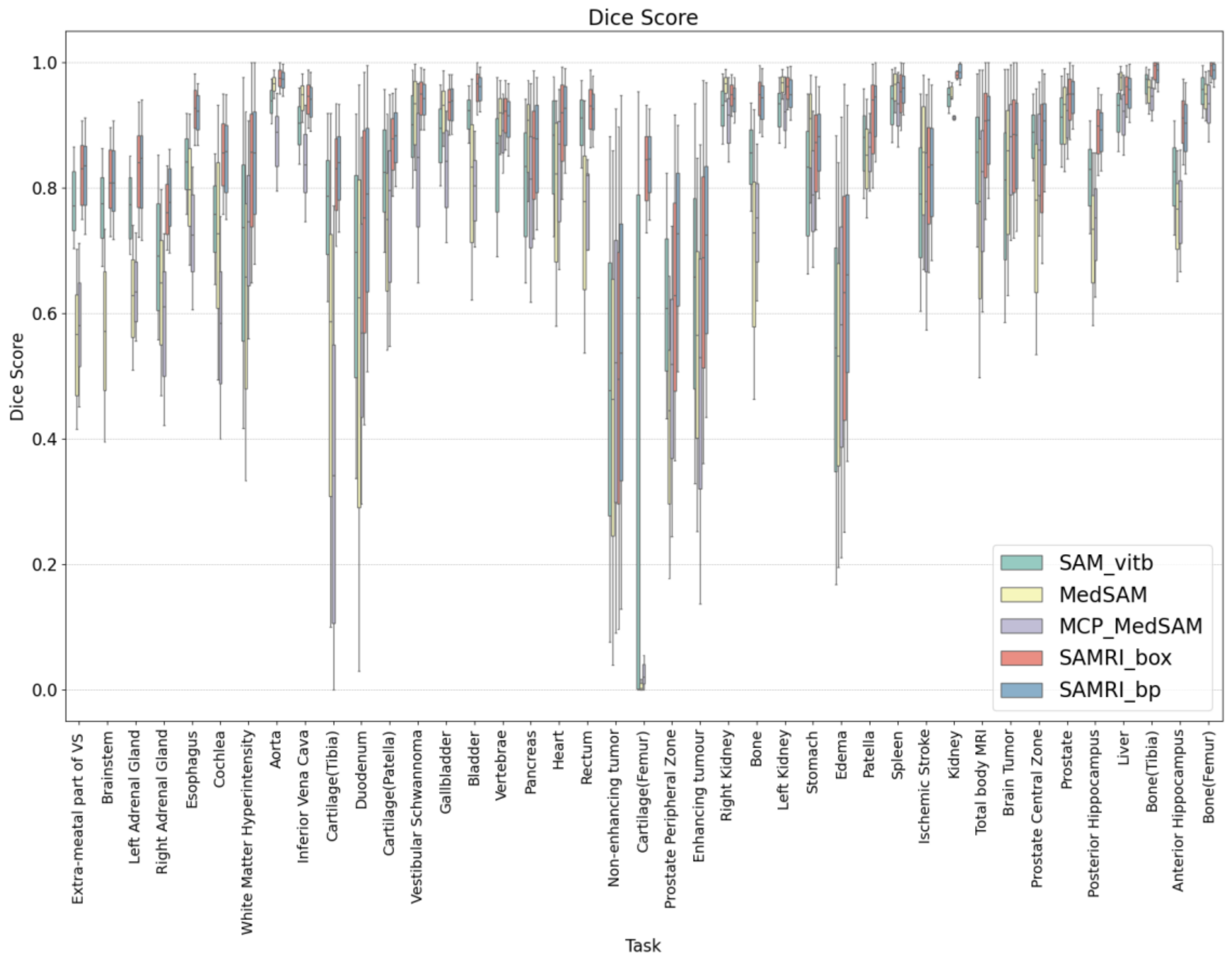

**Fig. S3.** *Boxplots of DSC across anatomical targets for all models.* The two versions of **SAMRI** shows consistently stronger performance on the majority of structures.

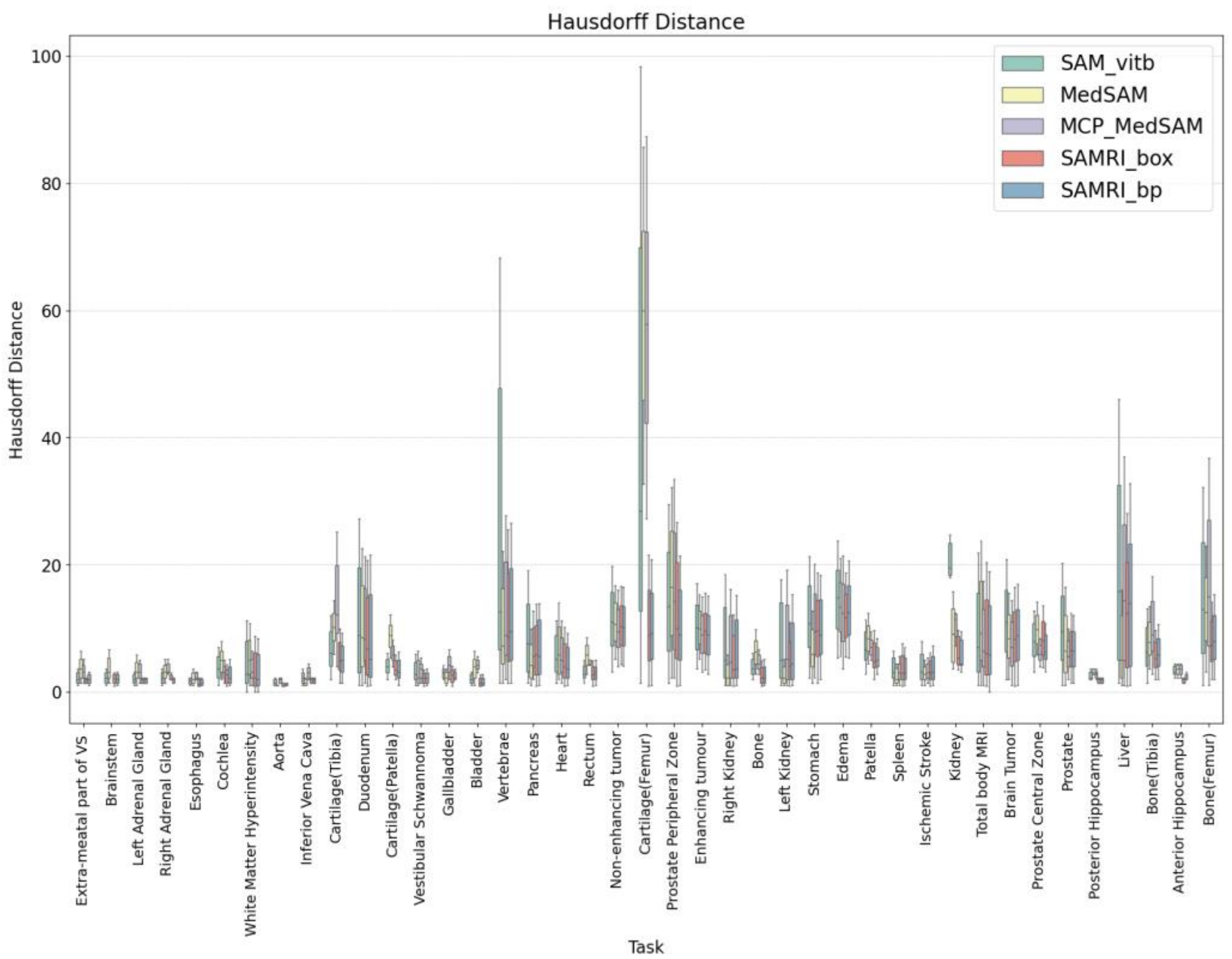

**Fig. S4**: *Boxplots of HD across anatomical targets for all models*.

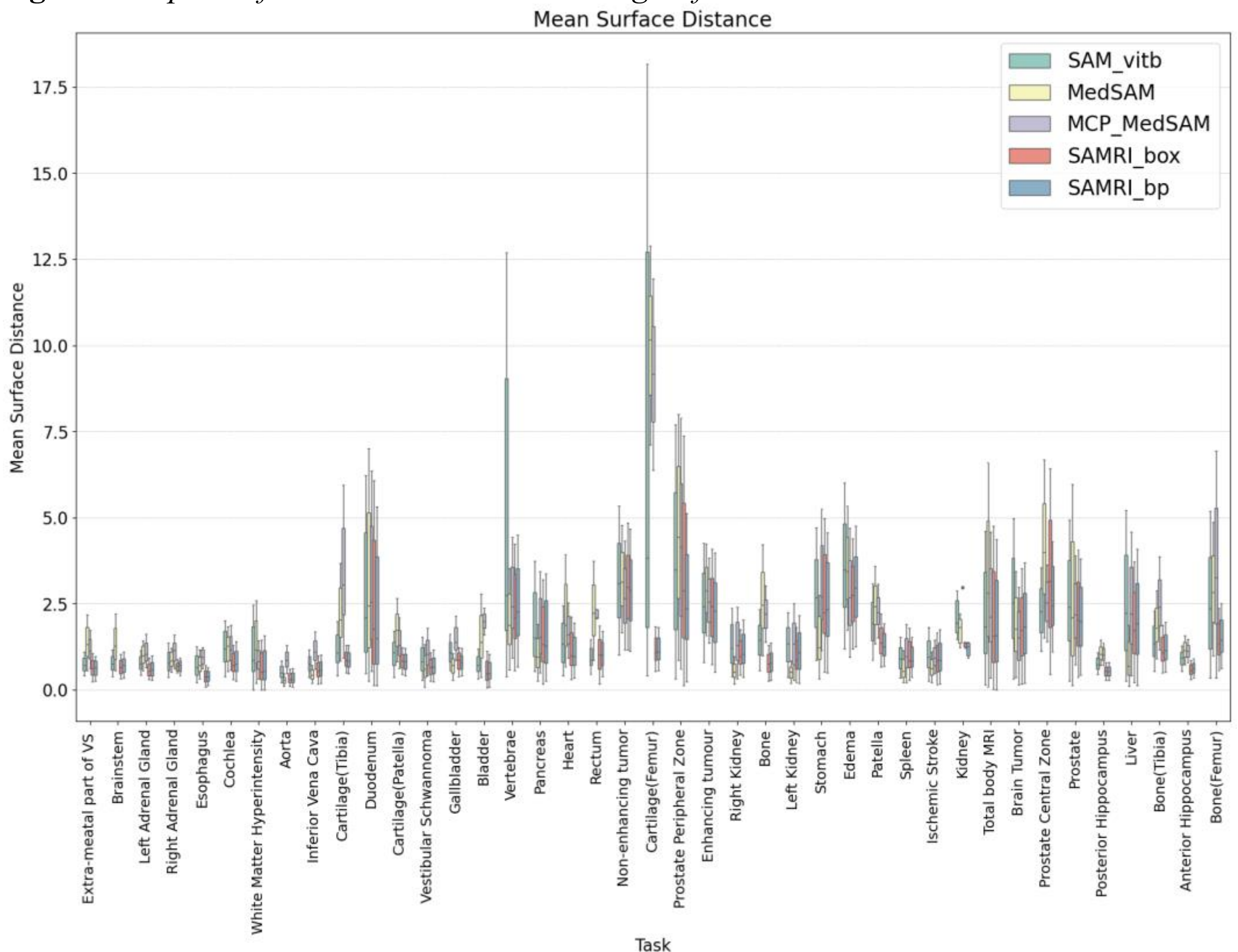

**Fig. S5**: *Boxplots of MSD across anatomical targets for all models*.

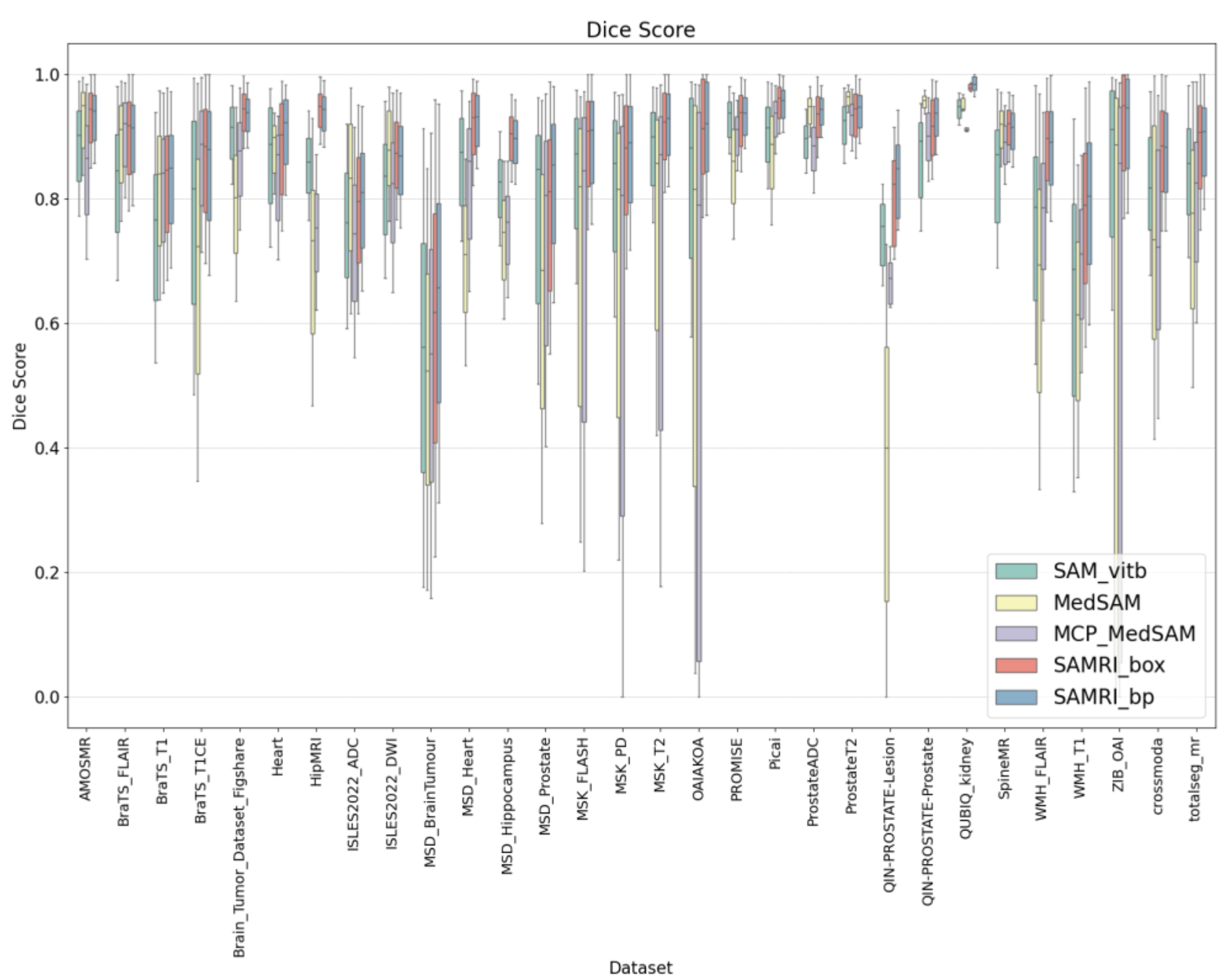

**Fig. S6**: *Boxplots of DSC across datasets for all models*. The two versions of **SAMRI** shows consistently stronger performance on the majority of structures.

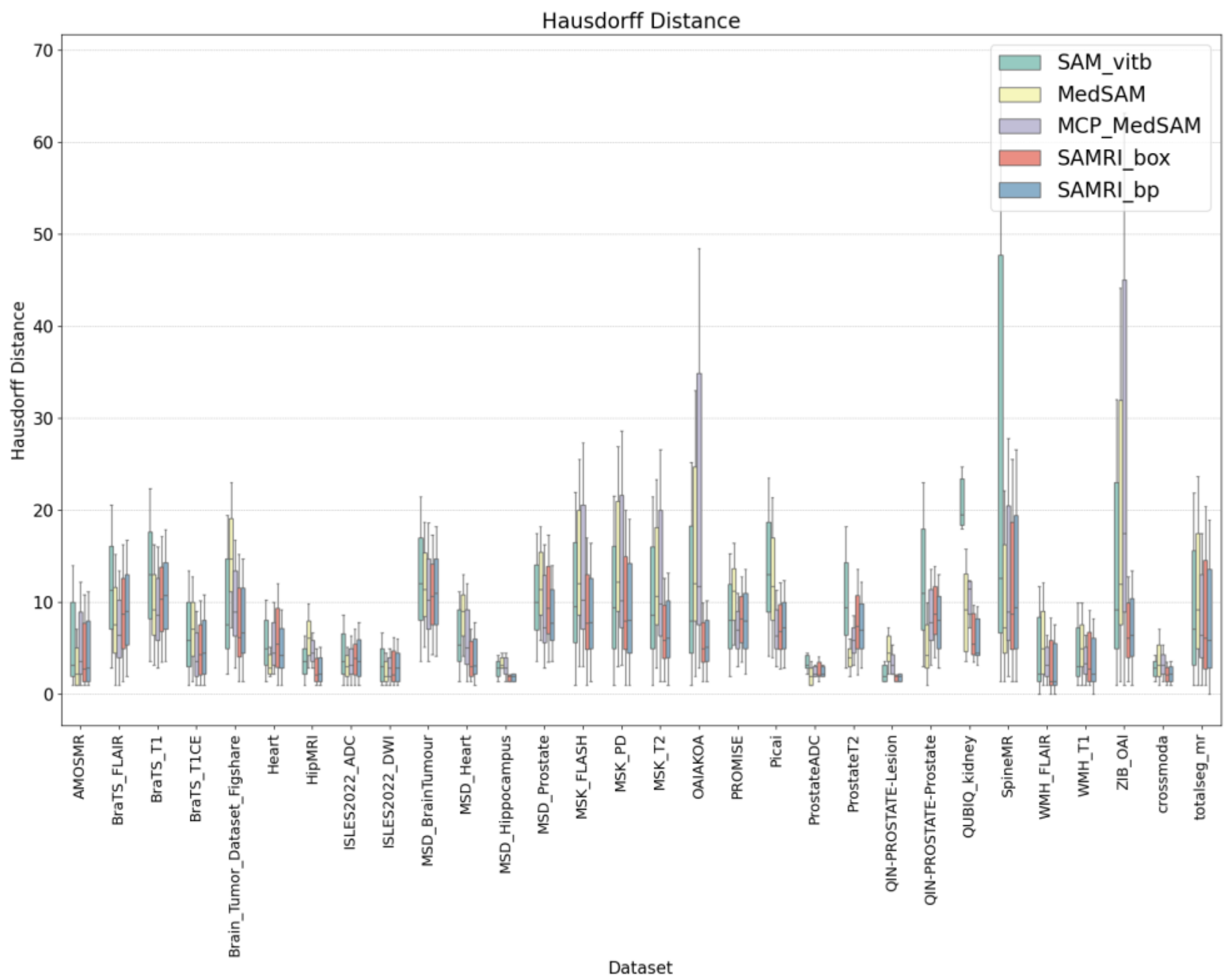

**Fig. S7**: *Segmentation performance on HD across datasets.*

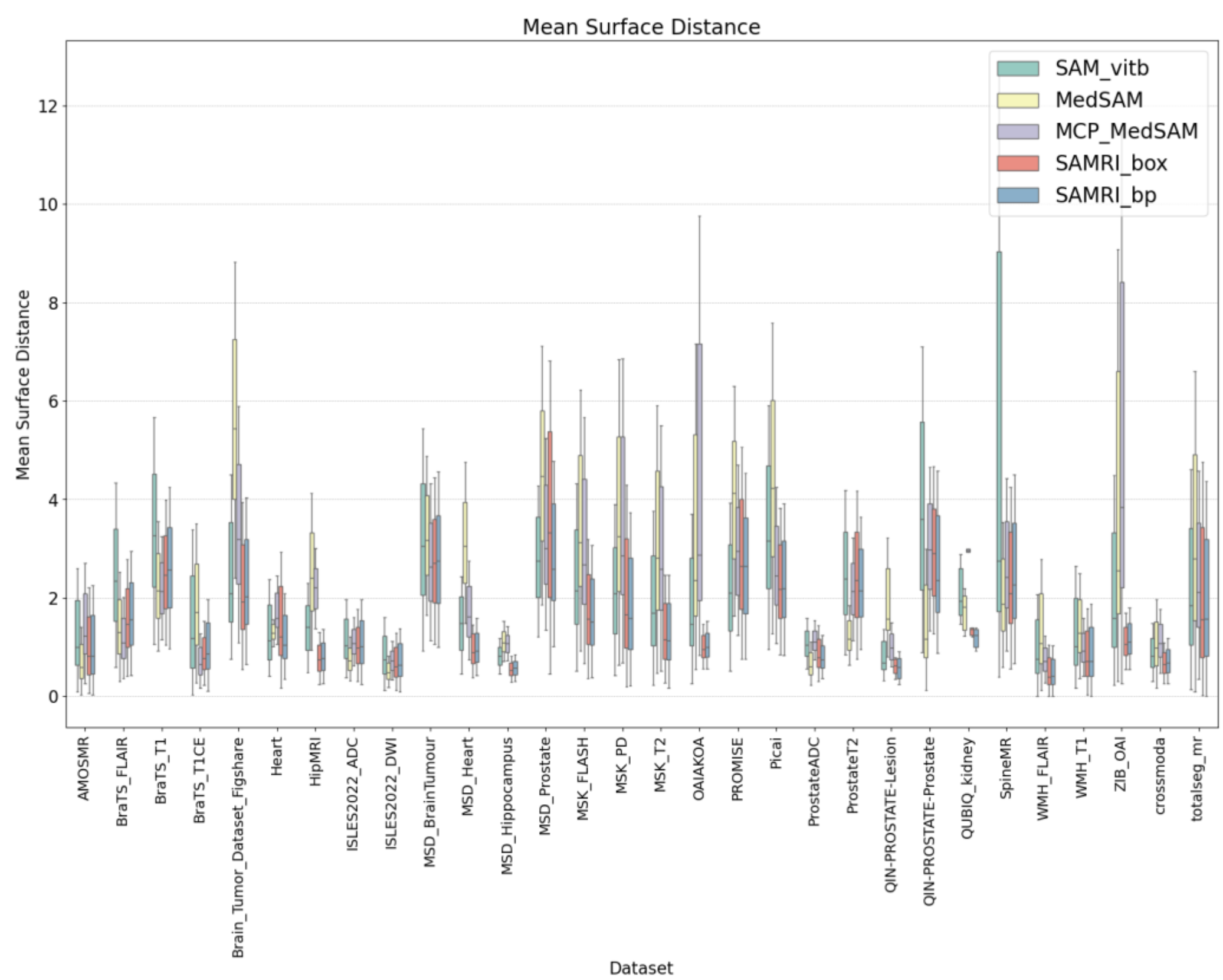

**Fig. S8**: *Segmentation performance on MSD across datasets.*

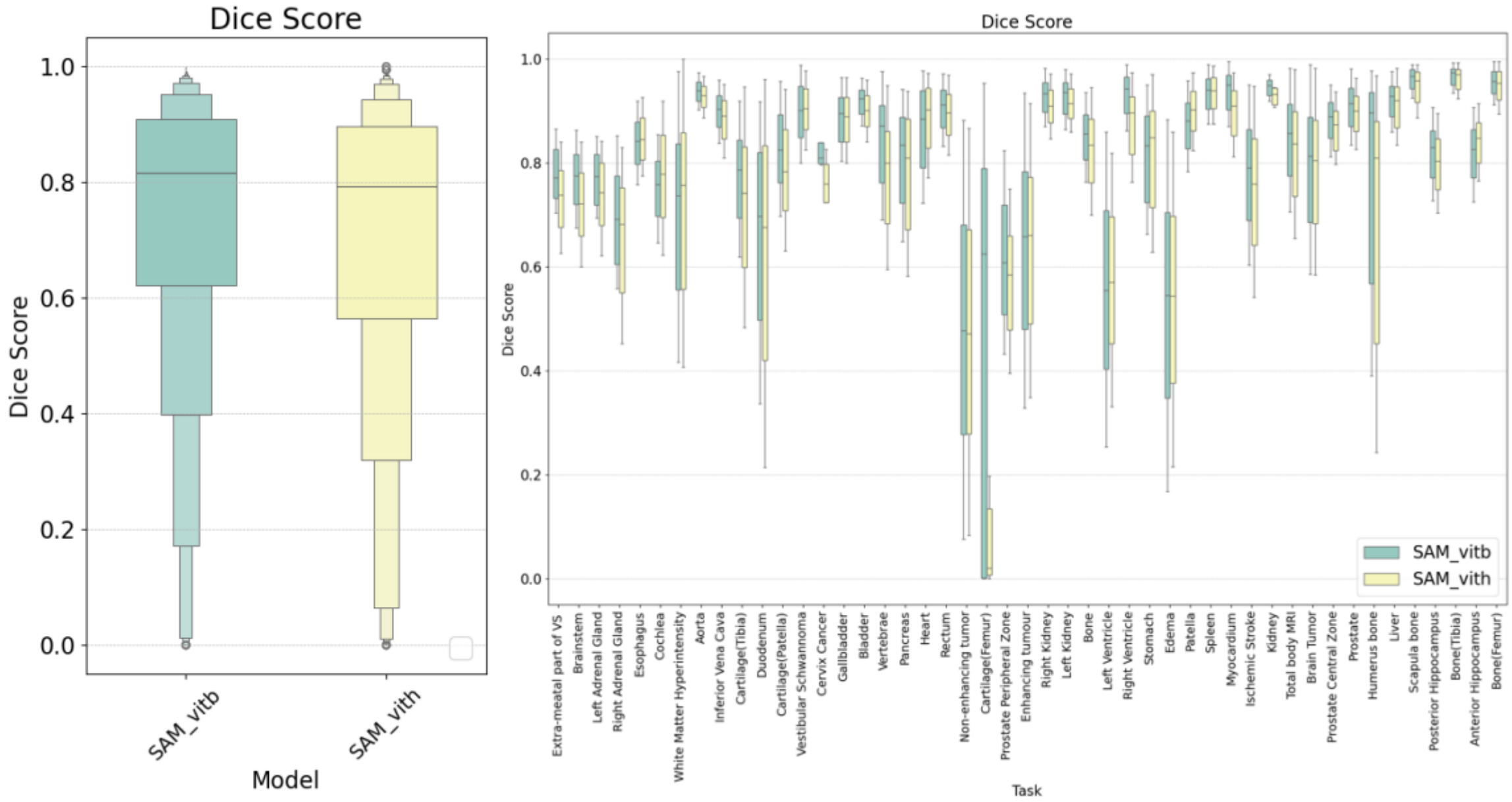

**Fig. S9** *Overall Dice distribution: SAM ViT-B vs. SAM ViT-H (zero-shot MRI).*
**Left:** Violin plots summarize Dice over all test cases using identical preprocessing and box-prompting. Medians and interquartile ranges are comparable, with **no consistent improvement from ViT-H** and a slightly lower central tendency in several runs. **Right:** Box plot of Dice across all datasets.

## Additional Comparative Model Evaluation

For completeness, we tested three additional SAM-based models—Med-SA [44], SAM-Med2D [76], and SAMed—using their publicly released pretrained checkpoints on the full SAMRI test set (Fig. S10). All three models produced near-zero Dice scores across the evaluation suite, confirming that a single pretrained checkpoint does not generalize across our diverse range of MRI contrasts, anatomies, and tasks. We attribute this to the mismatch between the narrow training distributions of these models (single anatomy, single modality, or limited MRI or other medical modality tasks) and the breadth of our evaluation. This outcome supports our inclusion criterion of evaluating only models with checkpoints suitable for broad, MRI-wide zero-shot evaluation.

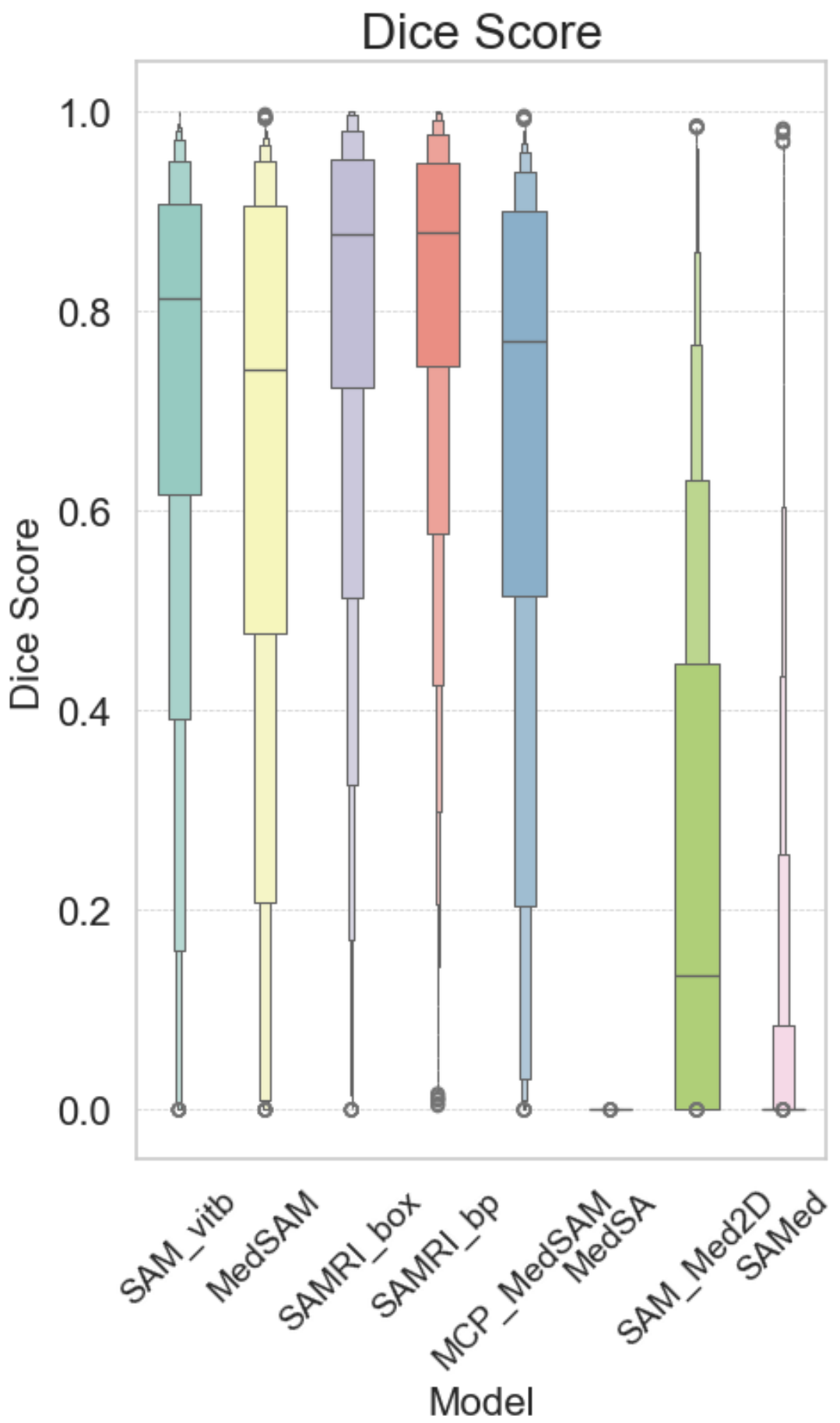

**Fig. S10**: Dice score distributions for additional SAM-based models tested on the SAMRI evaluation suite. Box plots show median, IQR, and full range. Med-SA, SAM-Med2D, and SAMed each produced near-zero Dice across the test set, consistent with the mismatch between their narrow training distributions and the breadth of our multi-anatomy, multi-contrast evaluation. SAMRI (box and box+point variants) and MedSAM are included for reference.

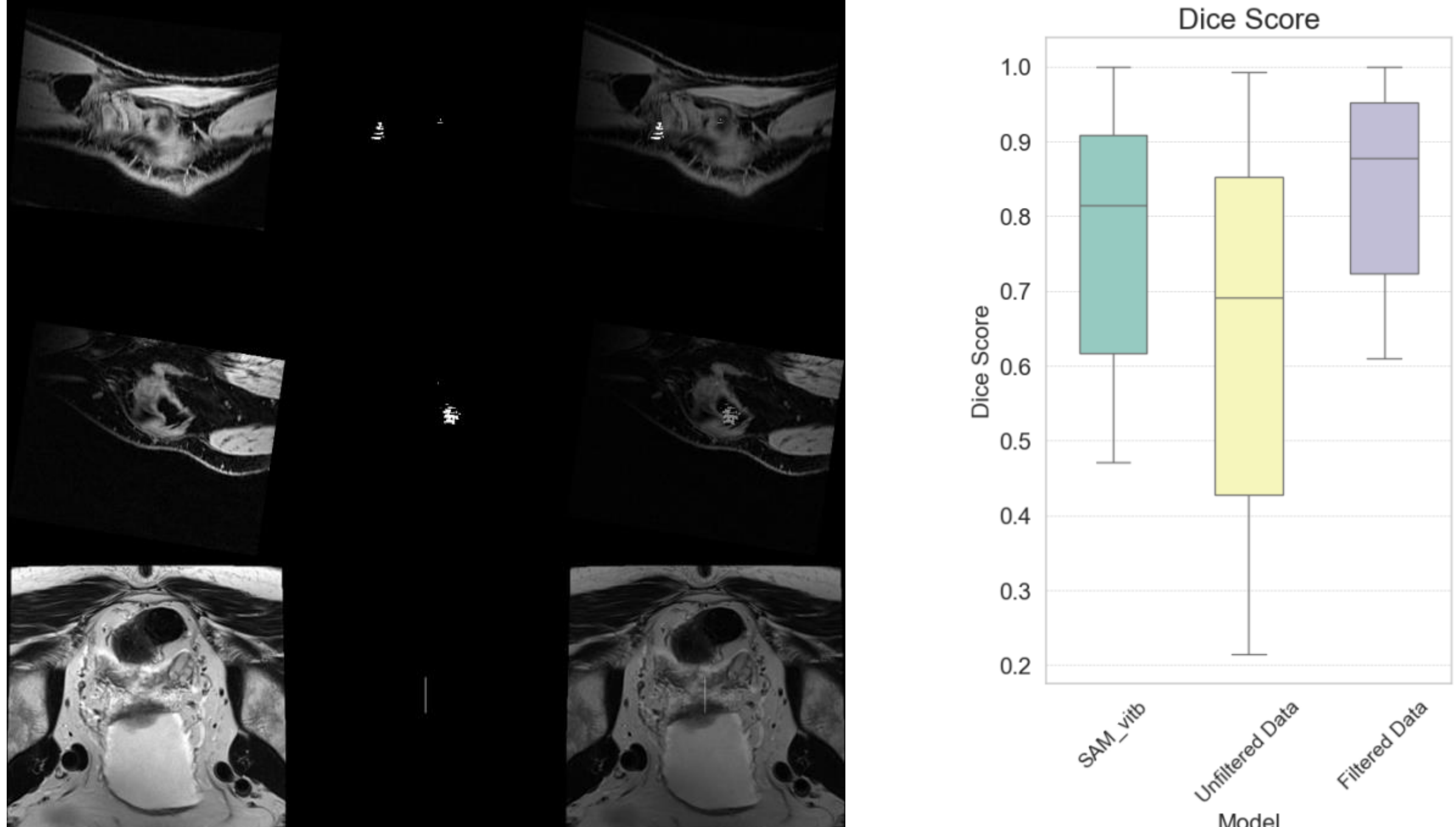

**Fig. S11:** *Effect of small-mask filtering on training quality and model performance.* **(Left)** Representative examples of masks occupying fewer than 10% of the largest slice's mask area within the same 3D volume; these arise from peripheral slices where the target structure is only partially visible, producing incomplete coverage or annotation artifacts. **(Right)** Dice score distributions comparing SAM ViT-B (zero-shot), SAMRI trained on unfiltered data (noisy masks included), and SAMRI trained on filtered data (masks below the 10% threshold removed); filtering consistently raises Dice scores and reduces variance, confirming that partial-structure masks degrade training stability. Validation and test sets are left unfiltered to preserve evaluation integrity.

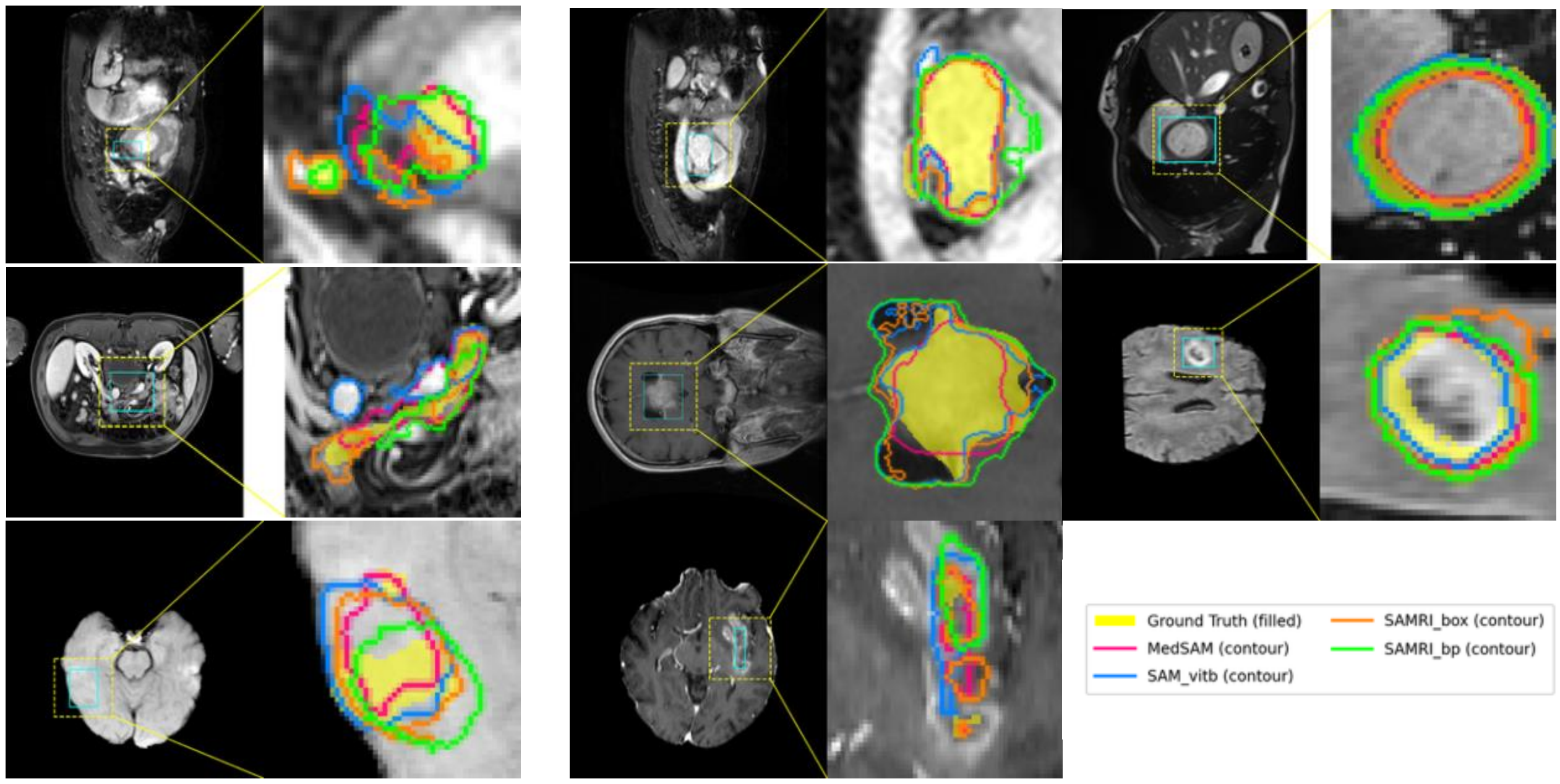

**Fig. S12**: *Underperformance examples.* Representative failure cases with full views and zoom-ins. Errors cluster in (i) low-SNR or blurred boundaries, (ii) thin/rim structures where small offsets cause large Dice

drops, (iii) multi sub-area object, (iv) strong neighboring edges that "pull" contours, and (v) very small targets and slices.

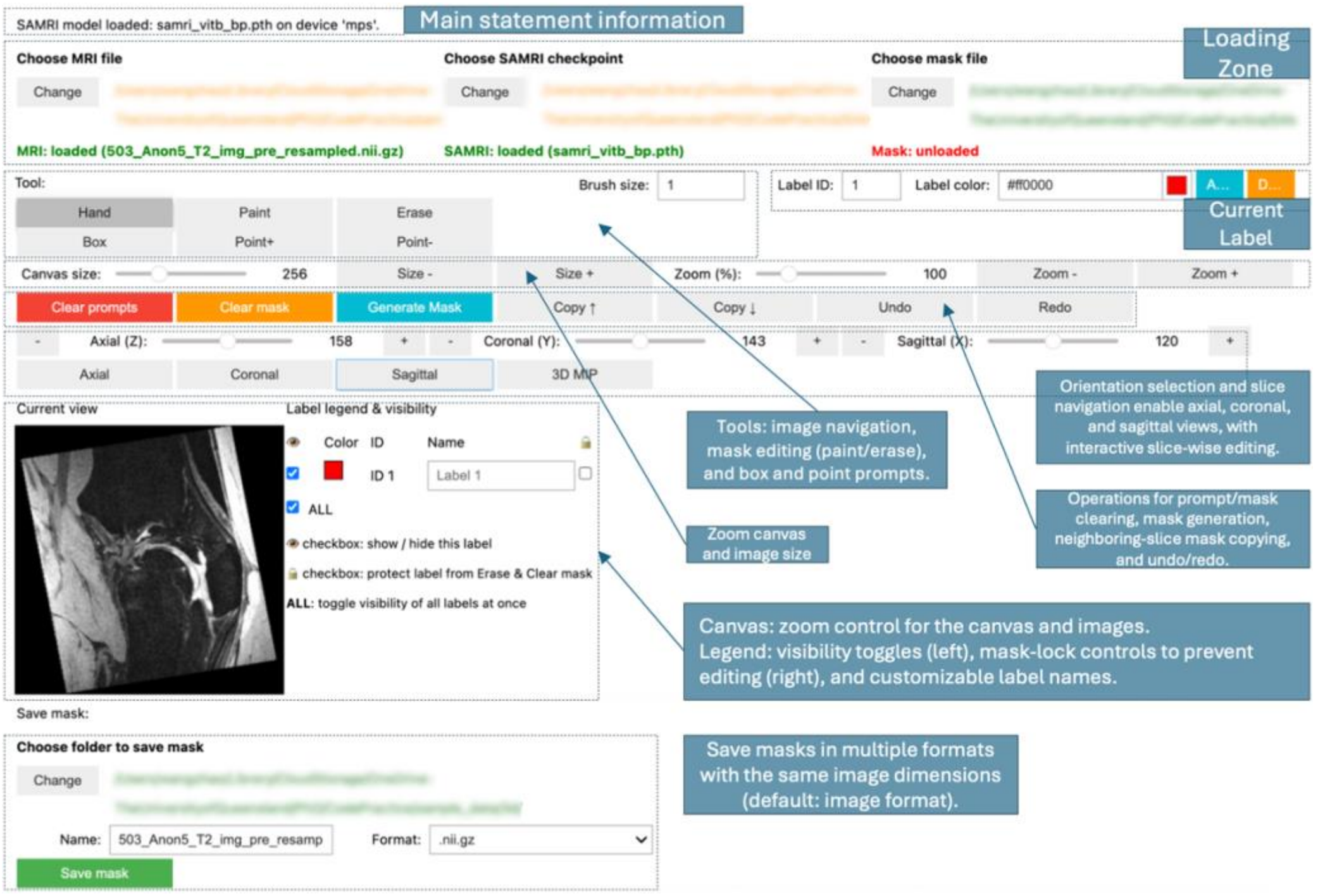

**Fig. S13**: *Practical deployment and interactive segmentation interface for SAMRI.* The local GUI supports interactive MRI visualization, prompt-based segmentation, and real-time mask refinement across axial, coronal, and sagittal views, with segmentation results saved locally in standard medical image formats.

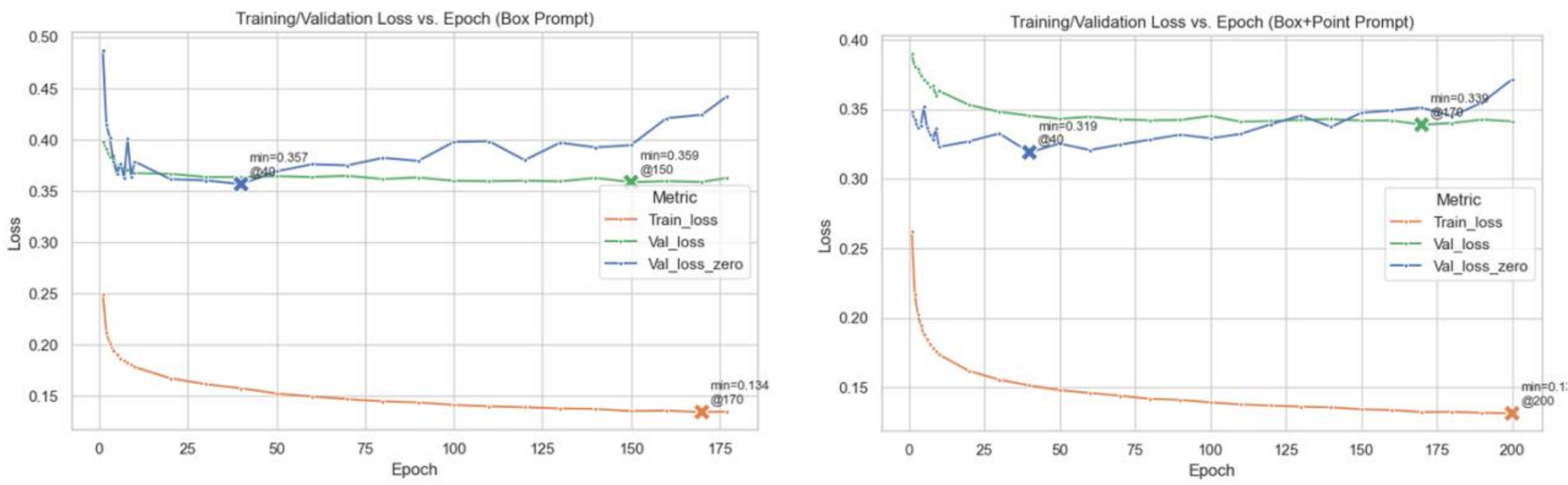

**Fig. S14**: *Loss trajectories across epochs.* Left: box prompt; right: box+point prompt. Plotted are training loss (orange), seen-set validation loss (green), and zero-shot validation loss (blue). The curves indicate stable convergence without overfitting on seen datasets, with optimal zero-shot performance at around epoch 40.

## Appendix S1: Training Efficiency Gain

Without precomputing embeddings, the wall-clock training time over N epochs is:

$$T_{total} = N \times (T_{data_preparing} + T_{encoder_infer} + T_{decoder_infer} + T_{back_updating})$$

Where $T_{total}$ is the total training time, $T_{data_preparing}$ is the data loading/preprocessing time, $T_{encoder_infer}$ is the image encoder inference time, $T_{decoder_infer}$ is the mask decoder inference time, and $T_{back_updating}$ is the backpropagation time.

With our two-stage pipeline (one-time embedding, then decoder-only training), the time becomes:

$$T_{pipeline} = T_{data_preparing} + T_{encoder_infer} + N \times (T_{decoder_infer} + T_{back_updating})$$

Because the encoder is removed from the per-epoch loop, we avoid (N-1) $T_{encoder_infer}$ of repeated computation. In our setting (N=100 epochs), eliminating those 99 encoder passes yielded an ~88% reduction in wall-clock training time.

## Appendix S2: Focal Loss and Dice Loss

**Focal Loss: [48]**Let S and G represent the predicted segmentation and ground truth mask, respectively, and let $s_i$ , $g_i$ denote the predicted and true values at pixel i across all N pixels in an image. The focal loss ($L_{Focal}$)is defined as:

$$L_{Focal} = -\frac{1}{N}\sum_{i=1}^{N} \alpha \cdot (1-p_i)^{\gamma} \cdot log(p_i)$$

Where:

- $p_i = s_i$ if $g_i = 1$, otherwise $p_i = 1 - s_i$
- $\alpha \in [0,1]$: weighting factor for balancing positive vs. negative samples (optional)
- $\gamma > 0$: focusing parameter that reduces the relative loss for well-classified examples
- $log(p_i)$: standard cross-entropy term

**Dice Loss: [49]** The Dice loss $L_{Dice}$ is defined as:

$$L_{Dice} = 1 - \frac{2\sum_{i=1}^{N} s_i g_i}{\sum_{i=1}^{N}(s_i)^2 + \sum_{i=1}^{N}(g_i)^2}$$

The final loss $L_{SAMRI}$ is defined as:

$$L_{SAMRI} = 20 \cdot L_{Focal} + L_{Dice}$$

This formulation is particularly effective for small object segmentation, as focal loss prevents rare pixels (often corresponding to small structures) from being overwhelmed during training, while Dice loss ensures accurate boundary delineation, which is crucial when objects are present in sparse, low-pixel-count regions.

## Appendix S3: Evaluation Metrics

We evaluate model performance using three complementary metrics that capture both region overlap and boundary accuracy:

**Dice Similarity Coefficient (DSC)** [50]: Measures region overlap between predicted segmentation S and ground truth G:

$$DSC(S,G)=\frac{2\sum_{i=1}^{N} s_i g_i}{\sum_{i=1}^{N} s_i + \sum_{i=1}^{N} g_i}$$

Where:

- N: total number of pixels in the image.
- $s_i \in set(0,\ 1)$: predicted value for the i-th pixel (1 for foreground, 0 for background).
- $g_i \in set(0,\ 1)$: ground truth label for the i-th pixel.

**Hausdorff distance (HD)**: Measures the maximum distance between predicted and ground truth surface points, capturing worst-case boundary errors:

$$d_H(A,B)=max[\sup_{a\in A} inf_{b\in B} d(a,b),\ _{b\in B} inf_{a\in A} d(b,a)]$$

Where A and B are the sets of surface points from the predicted and ground truth masks, respectively, and sup represents the supremum operator, $inf$ the infimum operator

**Mean surface distance (MSD)**: Provides the average bidirectional boundary error which is less sensitive to outliers than the HD:

$$MSD(A,B)=\frac{1}{2}\left(\frac{1}{|A|}\sum_{a\in A} d(a,B)+\frac{1}{|B|}\sum_{b\in B} d(b,A)\right)$$

Where:

- $|A|$ and $|B|$ are the number of points in sets A and B, respectively.
- $d(a,B)=min_{b\in B} d(a,b)$
- $d(b,A)=min_{a\in A} d(b,a)$

## References

[40] Ma J, He Y, Li F, Han L, You C, Wang B. Segment anything in medical images. Nat Commun. 2024;15(1):654.

[42] Dong G, Wang Z, Chen Y, Sun Y, Song H, Liu L, Cui H. An efficient segment anything model for the segmentation of medical images. Sci Rep. 2024;14(1):19425.

[43] Chen C, Miao J, Wu D, Zhong A, Yan Z, Kim S, Hu J, Liu Z, Sun L, Li X. Ma-sam: Modality-agnostic sam adaptation for 3d medical image segmentation. Med Image Anal. 2024;98:103310.

[44] Wu J, Wang Z, Hong M, Ji W, Fu H, Xu Y, Xu M, Jin Y. Medical SAM adapter: Adapting segment anything model for medical image segmentation. Med Image Anal. 2025;102:103547.

[45] Gao Y, Xia W, Hu D, Wang W, Gao X. Desam: Decoupled segment anything model for generalizable medical image segmentation. In: International Conference on Medical Image Computing and Computer-Assisted Intervention; 2024 (pp. 509–519).

[46] Lei W, Xu W, Li K, Zhang X, Zhang S. MedLSAM: Localize and segment anything model for 3D CT images. Med Image Anal. 2025;99:103370.

[47] Lyu D, Gao R, Staring M. MCP-MedSAM: A powerful lightweight medical segment anything model trained with a single GPU in just one day. Mach Learn Biomed Imaging. 2025;3:135–151. https://doi.org/10.59275/j.melba.2025-4849.

[48] Lin TY, Goyal P, Girshick R, He K, Dollár P. Focal loss for dense object detection. In: Proceedings of the IEEE international conference on computer vision (pp. 2980–2988).

[49] Sudre CH, Li W, Vercauteren T, Ourselin S, Jorge Cardoso M. Generalised dice overlap as a deep learning loss function for highly unbalanced segmentations. In: Deep Learning in Medical Image Analysis and Multimodal Learning for Clinical Decision Support: Third International Workshop, DLMIA 2017, and 7th International Workshop, ML-CDS 2017, Held in Conjunction with MICCAI 2017, Québec City, QC, Canada, September 14, Proceedings 3 (pp. 240–248).

[50] Dice LR. Measures of the amount of ecologic association between species. Ecology. 1945;26(3):297–302.

[51] Bernard O, Lalande A, Zotti C, Cervenansky F, Yang X, Heng PA, Cetin I, Lekadir K, Camara O, Ballester MAG. Deep learning techniques for automatic MRI cardiac multi-structures segmentation and diagnosis: is the problem solved?. IEEE Trans {Med Imaging. }2018;37(11):2514–2525.

[52] Ji Y, Bai H, Ge C, Yang J, Zhu Y, Zhang R, Li Z, Zhanng L, Ma W, Wan X. Amos: A large-scale abdominal multi-organ benchmark for versatile medical image segmentation. Adv Neural {Inf Process Syst. }2022;35:36722–36732.

[53] Cheng J, Huang W, Cao S, Yang R, Yang W, Yun Z, Wang Z, Feng Q. Enhanced performance of brain tumor classification via tumor region augmentation and partition. PLoS One. 2015;10(10):e0140381.

[54] Cheng J, Yang W, Huang M, Huang W, Jiang J, Zhou Y, Yang R, Zhao J, Feng Y, Feng Q. Retrieval of brain tumors by adaptive spatial pooling and fisher vector representation. PLoS One. 2016;11(6):e0157112.

[55] Bakas S, Akbari H, Sotiras A, Bilello M, Rozycki M, Kirby J, Freymann J, Farahani K, Davatzikos C. Segmentation labels and radiomic features for the pre-operative scans of the TCGA-GBM collection (2017). DOI: https://doi. org/10.7937 K. ;9.

[56] Bakas S, Akbari H, Sotiras A, Bilello M, Rozycki M, Kirby J, Freymann J, Farahani K, Davatzikos C. Segmentation labels and radiomic features for the pre-operative scans of the TCGA-LGG collection [Data Set]. The Cancer Imaging Archive. 2017.

[57] Bakas S, Akbari H, Sotiras A, Bilello M, Rozycki M, Kirby JS, Freymann JB, Farahani K, Davatzikos C. Advancing the cancer genome atlas glioma MRI collections with expert segmentation labels and radiomic features. Sci Data. 2017;4(1):1–13.

[58] Bakas S, Reyes M, Jakab A, Bauer S, Rempfler M, Crimi A, Shinohara RT, Berger C, Ha SM, Rozycki M. Identifying the best machine learning algorithms for brain tumor segmentation, progression assessment, and overall survival prediction in the BRATS challenge. 2018. arXiv preprint arXiv:1811.02629.

[59] Menze BH, Jakab A, Bauer S, Kalpathy-Cramer J, Farahani K, Kirby J, Burren Y, Porz N, Slotboom J, Wiest R. The multimodal brain tumor image segmentation benchmark (BRATS). IEEE Trans {Med Imaging. }2014;34(10):1993–2024.

[60] Bowen SR, Yuh WT, Hippe DS, Wu W, Partridge SC, Elias S, Jia G, Huang Z, Sandison GA, Nelson D. Tumor radiomic heterogeneity: Multiparametric functional imaging to characterize variability and predict response following cervical cancer radiation therapy. J Magn {Reson Imaging. }2018;47(5):1388–1396.

[61] Kavur AE, Gezer NS, Barış M, Aslan S, Conze PH, Groza V, Pham DD, Chatterjee S, Ernst P, Özkan S. CHAOS challenge-combined (CT-MR) healthy abdominal organ segmentation. Med Image Anal. 2021;69:101950.

[62] Dorent R, Kujawa A, Ivory M, Bakas S, Rieke N, Joutard S, Glocker B, Cardoso J, Modat M, Batmanghelich K. CrossMoDA 2021 challenge: Benchmark of cross-modality domain adaptation techniques for vestibular schwannoma and cochlea segmentation. Med Image Anal. 2023;83:102628.

[63] Pace DF, Contreras HTM, Romanowicz J, Ghelani S, Rahaman I, Zhang Y, Gao P, Jubair MI, Yeh T, Golland P, Geva T, Ghelani S, Powell AJ, Moghari MH. HVSMR-2.0: A 3D cardiovascular MR dataset for whole-heart segmentation in congenital heart disease. Sci Data. 2024;11(1):721.

[64] Hernandez Petzsche MR, de la Rosa E, Hanning U, Wiest R, Valenzuela W, Reyes M, Meyer M, Liew SL, Kofler F, Ezhov I. ISLES 2022: A multi-center magnetic resonance imaging stroke lesion segmentation dataset. Sci Data. 2022;9(1):762.

[65] Antonelli M, Reinke A, Bakas S, Farahani K, Kopp-Schneider A, Landman BA, Litjens G, Menze B, Ronneberger O, Summers RM, van Ginneken B, Bilello M, Bilic P, Christ PF, Do RKG, Gollub MJ, Heckers SH, Huisman H, Jarnagin WR, McHugo MK, Napel S, Pernicka JSG, Rhode K, Tobon-Gomez C, Vorontsov E, Meakin JA, Ourselin S, Wiesenfarth M, Arbeláez P, Bae B, Chen S, Daza L, Feng J, He B, Isensee F, Ji Y, Jia F, Kim I, Maier-Hein K, Merhof D, Pai A, Park B, Perslev M, Rezaiifar R, Rippel O, Sarasua I, Shen W, Son J, Wachinger C, Wang L, Wang Y, Xia Y, Xu D, Xu Z, Zheng Y, Simpson AL, Maier-Hein L, Cardoso MJ. The Medical Segmentation Decathlon. Nat Commun. 2022;13(1):4128.

[66] Simpson AL, Antonelli M, Bakas S, Bilello M, Farahani K, Van Ginneken B, Kopp-Schneider A, Landman BA, Litjens G, Menze B. A large annotated medical image dataset for the development and evaluation of segmentation algorithms. 2019. arXiv preprint arXiv:1902.09063.

[67] Woo B, Engstrom C, Baresic W, Fripp J, Crozier S, Chandra SS. Automated anomaly-aware 3D segmentation of bones and cartilages in knee MR images from the Osteoarthritis Initiative. Med Image Anal. 2024;93:103089.

[68] Clark K, Vendt B, Smith K, Freymann J, Kirby J, Koppel P, Moore S, Phillips S, Maffitt D, Pringle M, Tarbox L, Prior F. The Cancer Imaging Archive (TCIA): Maintaining and Operating a Public Information Repository. J Digit Imaging. 2013;26(6):1045–1057.

[69] Litjens G, Toth R, Van De Ven W, Hoeks C, Kerkstra S, Van Ginneken B, Vincent G, Guillard G, Birbeck N, Zhang J. Evaluation of prostate segmentation algorithms for MRI: the PROMISE12 challenge. Med Image Anal. 2014;18(2):359–373.

[70] Saha A, Hosseinzadeh M, Huisman H. End-to-end prostate cancer detection in bpMRI via 3D CNNs: effects of attention mechanisms, clinical priori and decoupled false positive reduction. Med Image Anal. 2021;73:102155.

[71] Fedorov A, Schwier M, Clunie D, Herz C, Pieper S, Kikinis R, Tempany C, Fennessy F. An annotated test-retest collection of prostate multiparametric MRI. Sci Data. 2018;5(1):1–13.

[72] Becker AS, Chaitanya K, Schawkat K, Muehlematter UJ, Hötker AM, Konukoglu E, Donati OF. Variability of manual segmentation of the prostate in axial T2-weighted MRI: a multi-reader study. Eur J Radiol. 2019;121:108716.

[73] Zukić D, Vlasák A, Egger J, Hořínek D, Nimsky C, Kolb A. Robust detection and segmentation for diagnosis of vertebral diseases using routine MR images. In: Computer Graphics Forum (pp. 190–204).

[74] D'Antonoli TA, Berger LK, Indrakanti AK, Vishwanathan N, Weiss J, Jung M, Berkarda Z, Rau A, Reisert M, Küstner T. Totalsegmentator mri: Robust sequence-independent segmentation of multiple anatomic structures in mri. Radiology. 2025;314(2):e241613.

[75] Kuijf HJ, Biesbroek JM, De Bresser J, Heinen R, Andermatt S, Bento M, Berseth M, Belyaev M, Cardoso MJ, Casamitjana A. Standardized assessment of automatic segmentation of white matter hyperintensities and results of the WMH segmentation challenge. IEEE Trans {Med Imaging. }2019;38(11):2556–2568.

[76] Sun J, Chen K, He Z, Ren S, He X, Liu X, Peng C. Medical image analysis using improved SAM-Med2D: segmentation and classification perspectives. BMC Med Imaging. 2024;24(1):241.